\documentclass[twocolumn]{aastex62}
\usepackage{natbib}
\graphicspath{{./}{../pictures/}}

\received{}
\revised{}
\accepted{}
\submitjournal{ApJ}

\shorttitle{{\sc TDEmass}}
\shortauthors{Ryu et al.}
\usepackage{newtxtext,newtxmath}
\usepackage{natbib, amsmath, color}
\usepackage{graphicx, hyperref}

\usepackage[T1]{fontenc}

\DeclareRobustCommand{\VAN}[3]{#2}
\let\VANthebibliography\thebibliography
\def\thebibliography{\DeclareRobustCommand{\VAN}[3]{##3}\VANthebibliography}

\usepackage{graphicx}	
\usepackage{amsmath}	
\usepackage{amssymb}	
\usepackage{xcolor,ulem}
\usepackage{cases}
\usepackage{array,multirow}
\usepackage{upgreek}
\usepackage{textcomp}

\newcommand{\harm}{{\sc Harm3d}}   
\newcommand{\mesa}{{\small MESA}}

\newcommand{\erg}{{\rm erg}}

\newcommand{\s}{{\rm s}}
\newcommand{\cm}{~{\rm cm}}
\newcommand{\km}{~{\rm km}}
\newcommand{\g}{~{\rm g}}

\newcommand{\Lobs}{L_{\rm obs}}
\newcommand{\Tobs}{T_{\rm obs}}

\newcommand{\mbh}{M_{\rm BH}}
\newcommand{\mbhs}{M_{\rm BH,6}}
\newcommand{\mstar}{M_{\star}}
\newcommand{\rstar}{R_{\star}}
\newcommand{\Msun}{~M_{\odot}}
\newcommand{\Rsun}{~R_{\odot}}
\newcommand{\Kelvin}{~\mathrm{K}}

\defcitealias{Ryu1+2020}{Ryu2020a}
\defcitealias{Ryu2+2020}{Ryu2020b}
\defcitealias{Ryu3+2020}{Ryu2020c}
\defcitealias{Ryu4+2020}{Ryu2020d}

\defcitealias{Leloudas+2019}{1}
\defcitealias{Hinkle+2020}{2}
\defcitealias{Holoien2020}{3}
\defcitealias{Chornock+2013}{4}
\defcitealias{Gezari+2012}{5} 
\defcitealias{Nicholl+2019}{6}
\defcitealias{Holoien+2019}{7}
\defcitealias{Arcavi+2014}{8}
\defcitealias{Blagorodnova+2019}{9}
\defcitealias{velzen+2020}{10}

\defcitealias{McConnellMa2013}{I}
\defcitealias{Gultekin2009}{II}
\defcitealias{HaringRix2004}{III}
\defcitealias{KormendyHo2013}{IV}
\defcitealias{FerrareseFord2005}{V}

\begin{document}
	
	\title{Measuring stellar and black hole masses of tidal disruption events.}
	\correspondingauthor{Taeho Ryu}
	\email{tryu2@jhu.edu}
	
	\author[0000-0002-0786-7307]{Taeho Ryu}
	\affil{Physics and Astronomy Department, Johns Hopkins University, Baltimore, MD 21218, USA}
	
	\author{Julian Krolik}
	\affiliation{Physics and Astronomy Department, Johns Hopkins University, Baltimore, MD 21218, USA}
	\author{Tsvi Piran}
	\affiliation{Racah Institute of Physics, Hebrew University, Jerusalem 91904, Israel}

\begin{abstract}
The flare produced when a star is tidally disrupted by a supermassive black hole holds potential as a diagnostic of both the black hole mass and the star mass.  We propose a new method to realize this potential based upon a physical model of optical/UV light production in which shocks near the apocenters of debris orbits dissipate orbital energy, which is then radiated from that region. Measurement of the optical/UV luminosity and color temperature at the peak of the flare leads directly to the two masses. The black hole mass depends mostly on the temperature observed at peak luminosity, while the mass of the disrupted star depends mostly on the peak luminosity.
We introduce {\sc TDEmass}, a method to infer the black hole and stellar masses given these two input quantities. Using {\sc TDEmass}, we find, for 21 well-measured events,  black hole masses between $5\times 10^5$ and $10^7 M_\odot$ and disrupted stars with initial masses between 0.6 and $13M_\odot$. 
An  open-source {\sc python}-based tool for {\sc TDEmass} is available at
\href{https://github.com/taehoryu/TDEmass.git}{https://github.com/taehoryu/TDEmass.git}.
\end{abstract}




\section{Introduction}

Among the most interesting questions concerning Tidal Disruption Events (TDEs) are the mass $\mbh$ of the disrupting supermassive black hole and the mass $\mstar$ of the disrupted star. Knowledge of these masses is clearly essential for understanding and modeling the event.  A statistical sample of these values would enable us to understand better the population of stars around galactic centers and at the same time provide an alternative method to establish the masses of these black holes. Although stellar kinematics can be used to measure larger black hole masses, it is difficult to do so for the mass range associated with the most common galaxies.

Nearly all previous efforts to estimate the stellar and black hole masses associated with a TDE have done so using {\sc Mosfit} \citep{Mockler+2019} to fit a multi-parameter phenomenological model to the optical/UV lightcurve.
{\sc Mosfit} assumes rapid ``circularization" of the debris stream and formation of a small accretion disk that powers the event. Recently, several competing methods have emerged.  One, also assuming efficient circularization, fits the X-ray spectrum to a slim disk model in order to find the black hole mass and spin \citep{Wen+2020}.
However, the pace of circularization is currently a matter of debate and may be much slower than previously thought (e.g., as reviewed by \citealt{BonnerotStone2020}). Assuming slow circularization, \citet{Zhou+2020} proposed a method to infer the black hole mass and the disrupted star mass if the flare is powered by accretion of matter on highly eccentric orbits.

Here we provide a  different parameter-inference method appropriate to slow circularization.
This new method, {\sc TDEmass}, rests upon a physical model \citep{Piran+2015} in which the optical/UV emission originates in the outer shocks that form due to intersections of the debris streams  near their orbital apocenters. 
An early version of this model that was motivated by the numerical simulation of \cite{Shiokawa+2015}
was applied successfully to ASASSN-14li \citep{Krolik+2016}, showing consistency with the optical/UV luminosity as well as with the X-ray and radio emission. 
In that simulation, the stellar debris were not ``circularized". Instead, as the stellar debris returned to the black hole it formed a large extended flattened structure (an elliptical disk). The observed optical/UV luminosity is then powered by shocks that dissipate the debris' kinetic energy.

For fixed $M_{\rm BH}$, the observed peak luminosity, temperature and time scale of a TDE depend only on $\mstar$ and $\Delta E$, the width of the energy distribution of the bound debris.  The energy available is proportional to $\mstar$; $\Delta E$ determines both the time scale of the mass return time $t_0$, the orbital period at energy $-\Delta E$, and $a_0$, the apocenter distance for that energy and nearly zero angular momentum. The apocenter also sets the characteristic length scale at which the returning streams dissipate their kinetic energy and then radiate it.

We have recently completed a study of TDEs incorporating both realistic main-sequence internal stellar structure and full general relativity \citep{Ryu1+2020,Ryu2+2020,Ryu3+2020,Ryu4+2020}. 
In this work, we determined two correction factors, $\Psi$ and $\Xi$, which correct
traditional order-of-magnitude estimates. The first, $\Psi$, relates the physical tidal radius, the maximum orbital pericenter such that the star is totally disrupted, to the ``tidal radius" $r_{\rm t}  \equiv (\mbh/\mstar)^{1/3} \rstar $, where $\rstar$ is the stellar radius.  Although $\Psi$ is an important quantity for determining event rates, it plays no role in determining observable features of an individual event. The second, $\Xi$, relates the real spread in specific energy $\Delta E$ to the corresponding order of magnitude estimate $\Delta \epsilon = G \mbh \rstar/r_{\rm t}^2$ and plays a crucial role in the present work. 

Building upon these results, in this paper we will show how $\mstar$ and $\mbh$ can be inferred from the peak luminosity and temperature of a TDE. In \S \ref{sec:correction} we describe the correction factor $\Xi$.  We then turn to our physical model and explain how $\mbh$ and $\mstar$ determine observables in \S \ref{sec:model}.  Inverting these equations in \S \ref{sec:from_obs_to_par}, we show how to obtain $\mbh$ and $\mstar$ from the observations. We then employ our method to estimate the masses of a sample of TDEs in \S \ref{sec:results}. We discuss implications of our method and possible future extensions to improve it in \S~\ref{sec:discussion}.
We conclude and summarize our results in \S~\ref{sec:conclusions}.

\section{Debris Energy For Main Sequence Stars}
\label{sec:correction}

\begin{figure}
		\includegraphics[width=8.5cm]{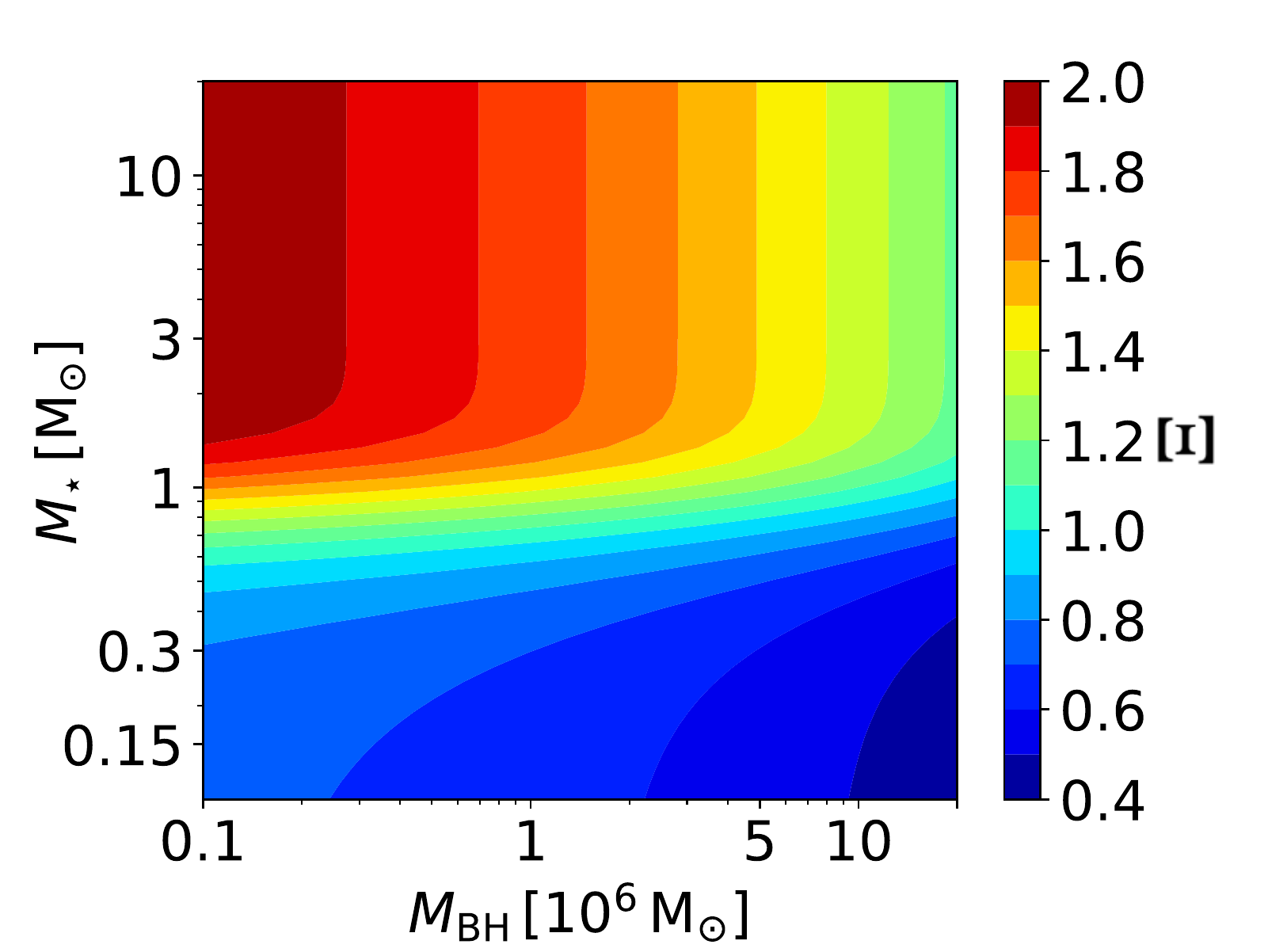}	
	\caption{The correction factor $\Xi\equiv \Delta E/ \Delta \epsilon$. }
	\label{fig:xi}
\end{figure}

In \citetalias{Ryu2+2020}, we simulated the disruption events of main-sequence stars  using hydrodynamics simulations in full general relativity (\harm, \citealt{Noble+2009}) and solving the Poisson equation for the stars' self-gravity in a relativistically consistent fashion.
The initial structures of stars spanning a wide range of mass $\mstar$ ($0.15-10\Msun$) were taken from \mesa ~\citep{Paxton+2011} evolutions to stellar middle-age, and their disruptions were studied for black hole masses over an even wider range ($10^5 - 5 \times 10^7 M_\odot$).  These improvements (exact relativistic tidal stresses with accurate self-gravity, realistic internal structure and wide ranges of $\mbh$ and $\mstar$) permitted quantitative determination of the outcome with significantly greater realism.  Although other parameters (e.g., stellar age, orbital pericenter) may also affect $\Delta E$, they do so much more weakly than $\mstar$ and $\mbh$ 
\citep{Law-Smith2019,Tejeda+2017,Gafton2019}.
Throughout the rest of this paper, $\mstar$ and $M_{\rm BH}$ will be given in units of $M_\odot$.

The width $\Delta E$ of the debris' specific energy distribution plays a key role in determining quantitative properties of the resulting flare.
$\Delta E$ can be factored into three parts \citep{Ryu1+2020}: the traditional order-of-magnitude estimate $\Delta\epsilon \equiv G\mbh\rstar/r_{\rm t}^2$; a function $\Xi_{\star}(\mstar$) describing the dependence arising from the star's internal structure:
\begin{align} \label{eq:Xi_star}
\Xi&_{\star}(M_{\star}) = \frac{0.62+\exp{[(M_{\star}-0.67)/0.21]}}{1 + 0.55~\exp{[(M_{\star}-0.67)/0.21]}} \ , 
\end{align}
and a function $\Xi_{\rm BH}$ describing additional $\mbh$-dependence due to relativistic effects:
\begin{align}\label{eq:Xi_BH}
\Xi&_{\rm BH}(M_{\rm BH}) =1.27 - 0.3\mbhs^{0.242},
\end{align}
where $\mbh = 10^{6}\mbhs$.
We define the product $\Xi \equiv \Xi_{\star} \Xi_{\rm BH} $, thus $\Delta E = \Xi ~\Delta\epsilon$.
While $\Xi$, shown in Figure~\ref{fig:xi},  is almost always within a factor of 2 of unity over  the span of $\mstar$ and $\mbh$ examined, its appearance at high powers in some of the expressions can make significant changes to the mass estimates.

\begin{figure}
\centering
		\includegraphics[width=4.5cm]{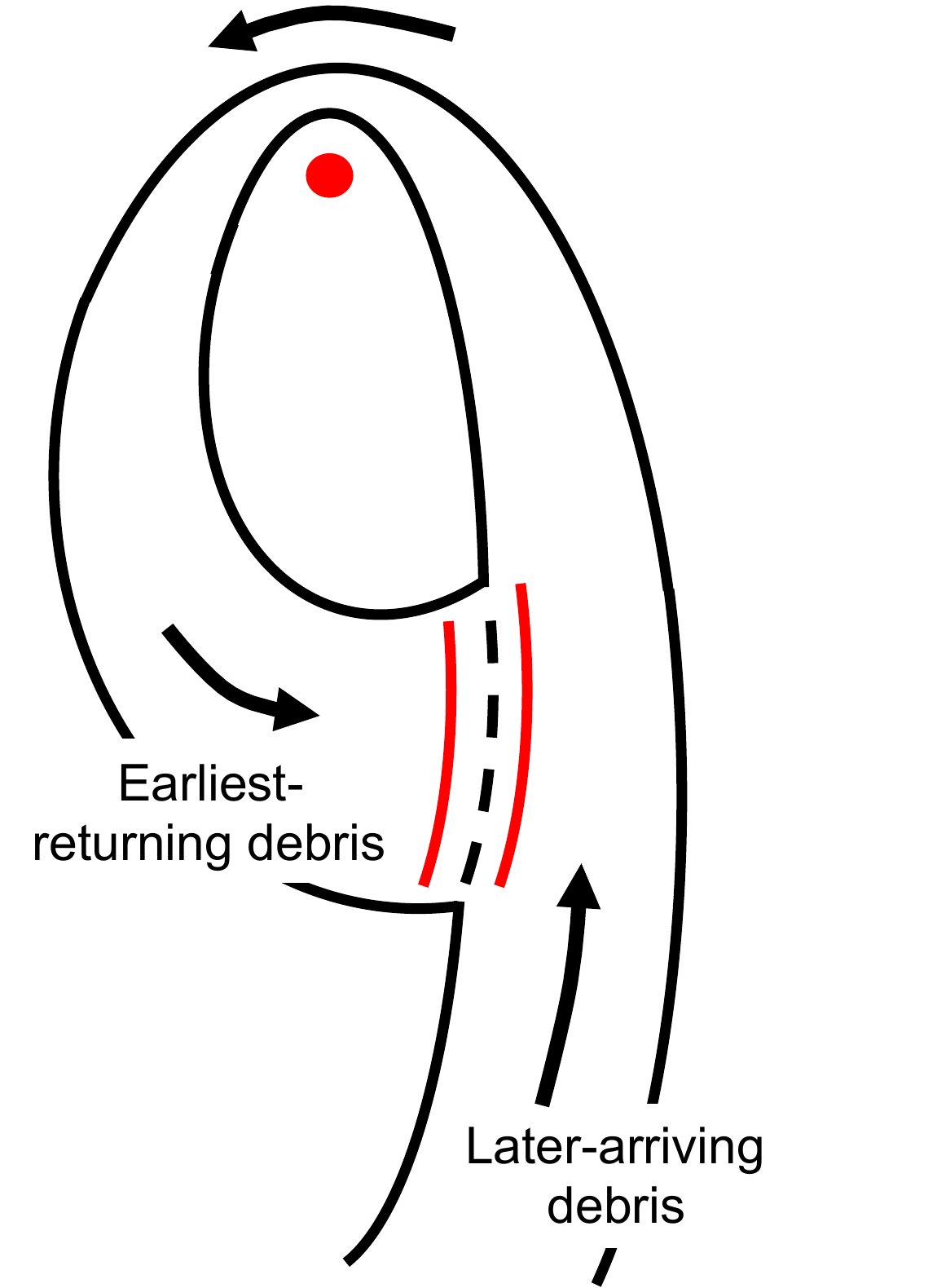}	
	\caption{Schematic picture illustrating the collisions between the earliest-returning debris and later-arriving debris near the apocenter after the second pericenter passage of the earliest-returning matter. The collisions lead to shocks (red solid lines) on both streams, which are separated by the contact discontinuity (black dashed line). The red circle near the top indicates the black hole. The black arrows show the flow of the streams.  Adapted from \citet{Shiokawa+2015}.}
	\label{fig:schematic}
\end{figure}

\section{The model: from parameters to observables}
\label{sec:model}

\citet{Piran+2015} proposed that the optical/UV light of tidal disruption events is powered by shocks  within an irregular, asymmetric, mildly-flattened, eccentric accretion flow formed from the bound debris. This model is based upon the results of a global hydrodynamical simulation of stellar debris dynamics that evolved the system until $\simeq 12~t_0$ after the tidal disruption \citep{Shiokawa+2015}. Here, we briefly summarize its key ingredients. When debris first falls back toward the black hole, it encounters matter that arrived earlier and has already passed once through the pericenter and traveled back out to near its apocenter,
as illustrated in Figure~\ref{fig:schematic}. The collision creates a pair of shocks with a contact discontinuity in between.
 The energy per newly-arriving mass dissipated in the shocks is $\approx (1/2)|\vec v_d - \vec v_o|^2$, where $\vec v_d$ is the debris velocity and $\vec v_o$ is the velocity of orbiting matter.  It is typically $\sim (1/2)(v_d^2 + v_o^2)$ because the angle between the two velocities is generally large (see Figure~\ref{fig:schematic}). If the shock occurs near the time of peak fallback rate, the specific dissipated energy is then $\sim \Delta E$.\footnote{ To be precise, when the fallback rate is near its peak, $(1/2)v_d^2 = \Delta E (2a/r -1)$, where $r$ is the distance from the black hole to the shock and $a$ is the semimajor axis.  For $r/a < 4/3$, $(1/2)v_d^2 > \Delta E/2$. Because $(1/2)v^2 = GM_{\rm BH}[1/r - 1/(2a)]$ and the material with $v_0$ has already suffered some energy loss, $(1/2)v_o^2$ should in general be slightly smaller than $(1/2)v_d^2$.   Their sum is therefore $\sim \Delta E$ for any collision point near, but not exactly at, the apocenter for highly-eccentric orbits with binding energy $\sim \Delta E$.}
 The bolometric luminosity at optical/UV wavelengths tracks the energy dissipation rate, which can be approximated as the product of the specific { dissipated energy} and the mass return rate.

 This model should be qualitatively valid so long as the apsidal precession angle of the debris stream upon returning to the pericenter is $< O(1)$. 
Large precession happens only for the small fraction of the events in which the disruption takes place at less than about 10 gravitational radii from the black hole \citep{Dai+15,Krolik+2020}.
Because the orbital energy loss in shocks near the apocenter is insufficient to circularize the tidal streams, and the formerly stellar matter has low specific angular momentum, the bound gas settles into an elliptical disk with a characteristic length scale $\sim a_{\rm 0}$, which is a factor $\sim (M_{\rm BH}/\mstar)^{1/3} \sim 100\times$ larger than the compact circular disk  (radius $\simeq 2r_{\rm t}$) often assumed to be the result of this process.

As demonstrated in \citet{Piran+2015}, this model predicts the characteristic scale of the peak luminosity, blackbody temperature, and line widths of TDEs. When extended to consider X-ray and radio observations, it also matches quite well the multiwavelength properties of an individual event, ASASSN 14li \citep{Krolik+2016}. However, these earlier efforts made cruder estimates of what we now call $\Xi$, making use of the correction factor for the energy width suggested by \citet{Phinney1989}. We improve their model by taking into account the $\Xi$ correction and then demonstrating how this model can be used for inferring $\mstar$ and $\mbh$ more generally.

As the typical energy of the bound material doesn't depend strongly on the star's pericenter provided it is greater than a few gravitational radii \citep{Tejeda+2017,Gafton2019,Ryu4+2020} and small enough to produce a full disruption, the system is characterized by three parameters, the black hole mass $\mbh$, the stellar mass $\mstar$ and the stellar radius $\rstar$. Adopting  a phenomenological $\mstar-\rstar$ relation \citep{Ryu2+2020}
\begin{align}
    \rstar=0.93\left(\frac{\mstar}{\Msun}\right)^{8/9} \Rsun \
\end{align} 
reduces this list to two.

\begin{figure}
	\centering
	\includegraphics[width=8.5cm]{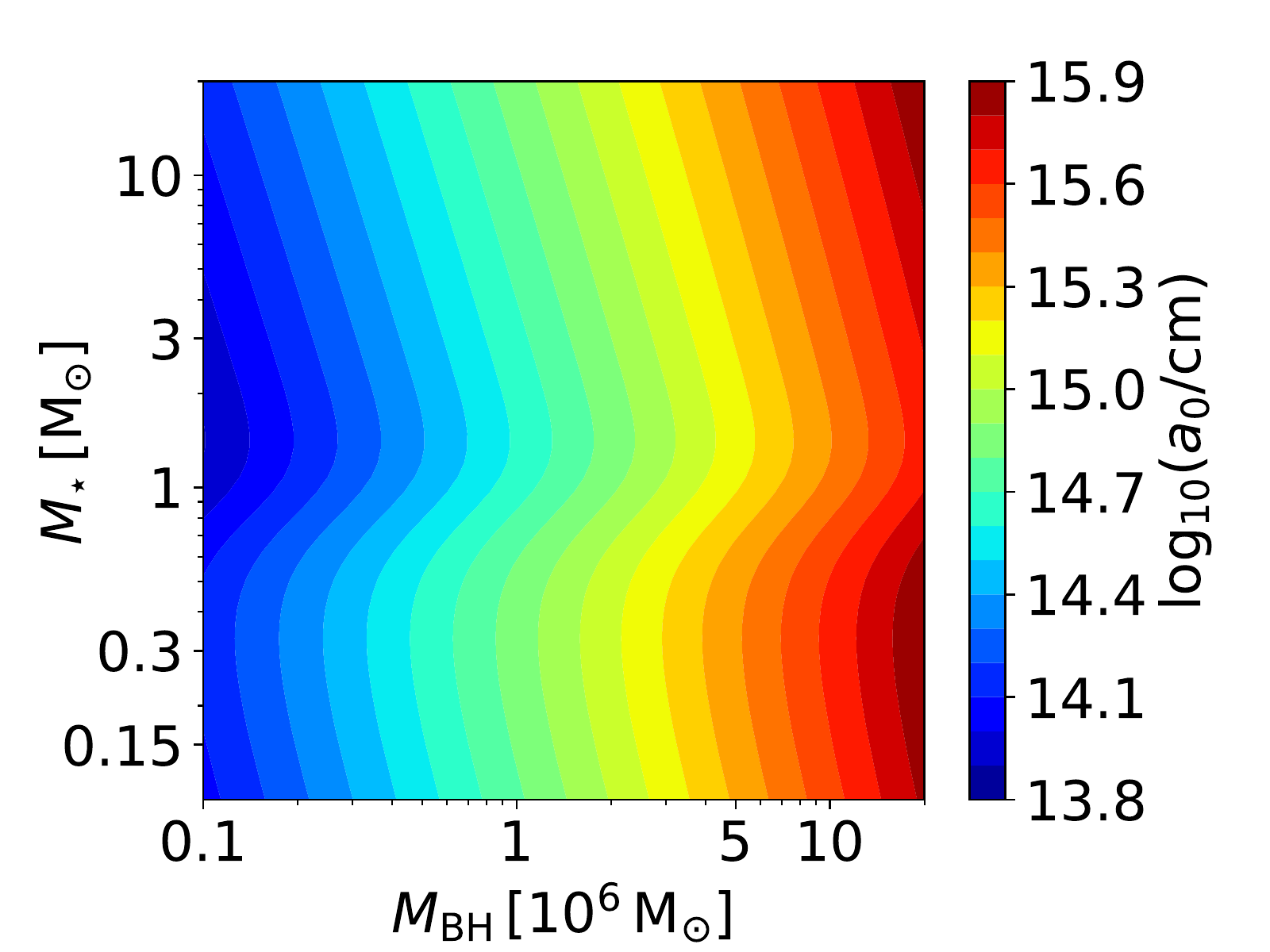}	
	\caption{The apocenter distance  $a_{0}$ in units of cm (Equation \ref{eq:a0}). 
	In our model $c_1 a_0$ is the size of the emitting region.}
	\label{fig:a0}
\end{figure}

The energy is produced by the infall of tidal streams to the previously described irregular accretion flow; we quantify the dissipation by supposing this flow has size $c_1 a_0$ and the shocks dissipate the associated free-fall kinetic energy.  
The period of peak mass fallback begins at $t_0$ after stellar pericenter passage.  Here
\begin{align} 
\label{eq:a0}
a_0&=
\frac{G \mbh}{\Delta E }=
6.5 \times 10^{14}~{\rm cm} ~ \mbhs^{2/3}   \mstar^{2/9}~ \Xi^{-1},
\end{align} 
and
\begin{figure}
	\centering
	\includegraphics[width=8.5cm]{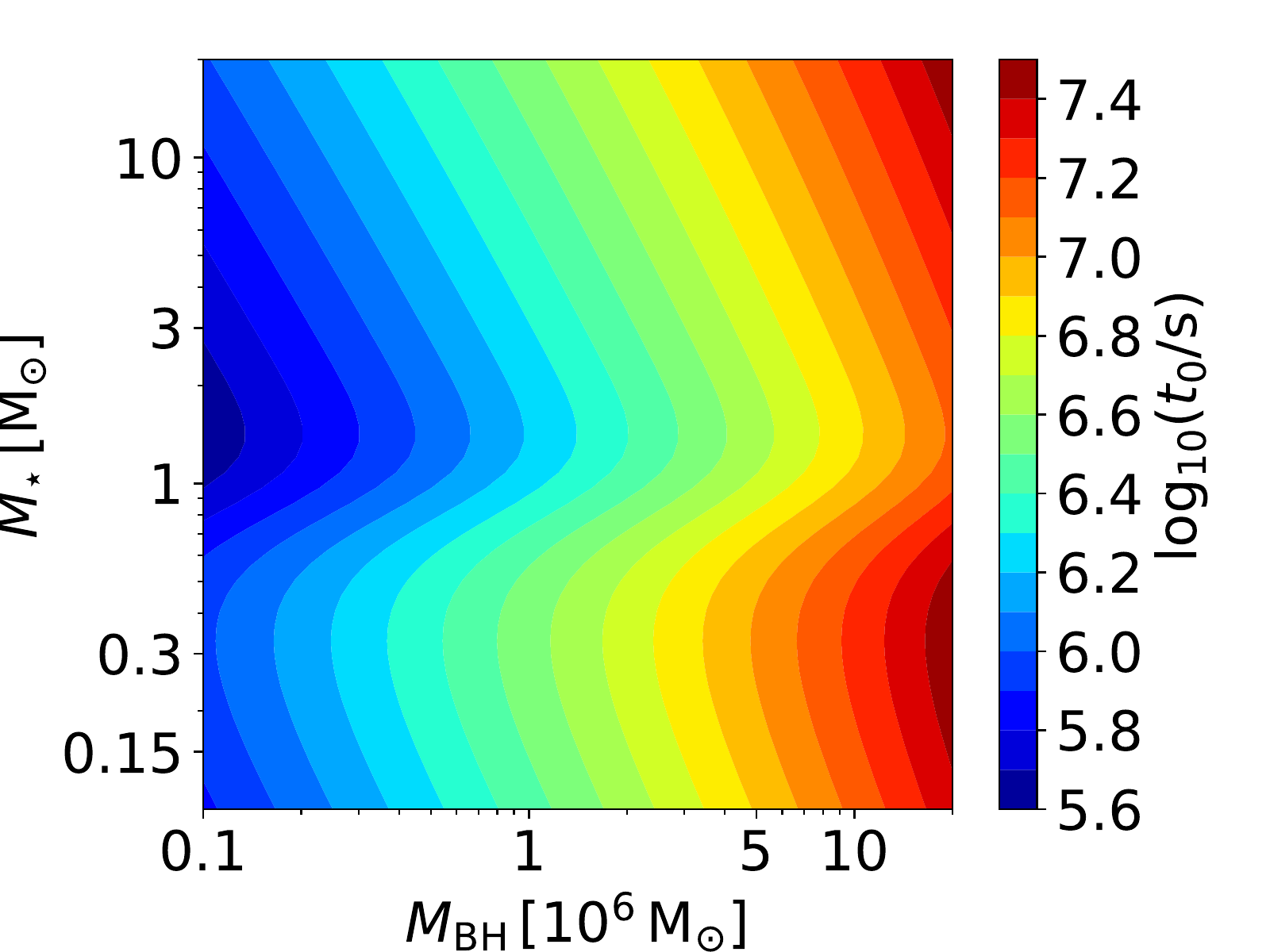}	
	\caption{The characteristic return time $t_{0}$ in units of seconds (Equation \ref{eq:t0}). }
	\label{fig:t0}
\end{figure}
\begin{align} \label{eq:t0}
t_{0}= \frac{\pi}{\sqrt{2}}  \frac{a_0^{3/2}}      {G^{1/2} \mbh^{1/2} } 
=3.2 \times 10^6 ~{\rm s}~ \mbhs^{1/2}  \mstar^{1/3} ~ \Xi^{-3/2}.
 \end{align}

\begin{figure}
    \includegraphics[width=8.5cm]{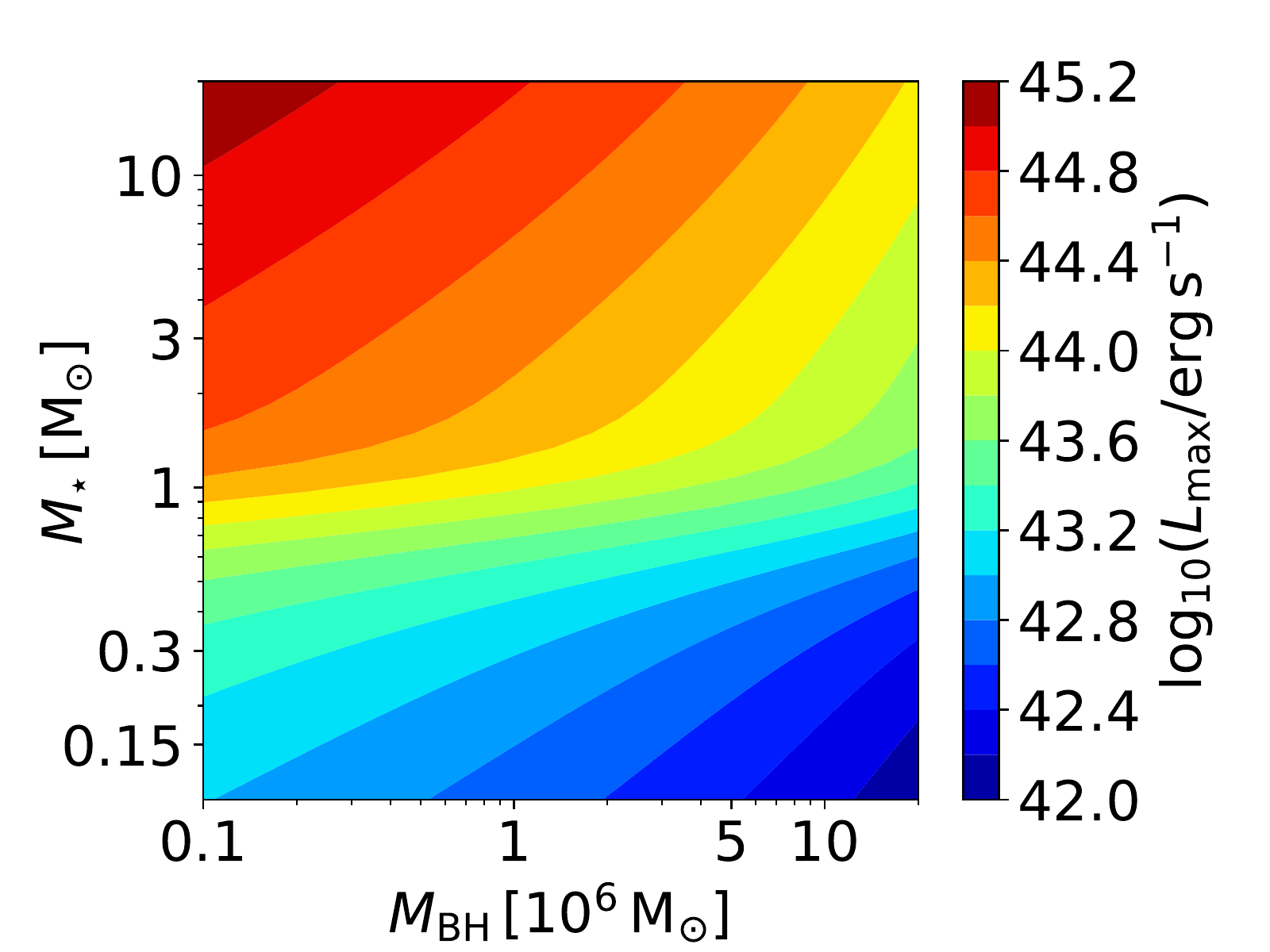}	
	\caption{The maximal luminosity $L_{\rm max}$ in units of  $\erg~ \s^{-1}$ for $c_{1}=1$.}
	\label{fig:lmax}
\end{figure}

 The apocenter distance $a_0$ is determined almost entirely by $M_{\rm BH}$ and is nearly independent of $\mstar$ (see Figure~\ref{fig:a0}).  This occurs because the $\mstar^{2/9}$ dependence (Equation~\ref{eq:a0}) is almost canceled by the gradual decrease of $\Xi^{-1}$ with $\mstar$.  Note that $c_1 a_0 (\Delta\Omega/4\pi)^{1/2}$ is equivalent to what is often called the ``blackbody radius".

Because $t_0$, which is shown in Figure~\ref{fig:t0}, is the orbital period for semimajor axis $a_0/2$, the shocks at $\sim a_0$ from the black hole begin at a time $1.5 t_0$ after the star passes pericenter, shortly after the peak mass fallback rate is reached.  At this time, the peak fallback rate is $\dot{M}_{\rm max}=\mstar/(3t_0)$ if the mass fallback rate post-peak is $\propto t^{-5/3}$, which is generally a good approximation for full disruptions.  Consequently, the maximal rate at which the outer shocks dissipate energy $L_{\rm max}$ is:
\begin{align} 
\label{eq:Lmax}
 L_{\rm max} &= \frac{G \mbh \dot{M}_{\rm max}}{c_{1}a_{0}}
= 4.3\times 10^{43}~c_{1}^{-1} \mstar^{4/9} \mbhs^{-1/6}~\Xi^{5/2}~{\rm erg/s}. 
\end{align}
We take $c_1=1$ as a fiducial value in the absence of more information. 
Figure \ref{fig:lmax} shows a contour plot for $L_{\rm max}$. 
As this figure clearly shows, $L_{\rm max}$ is more strongly dependent on $\mstar$ than on $M_{\rm BH}$.  In fact, the explicit dependence of $L_{\rm max}$ on $M_{\rm BH}$ is so weak, it depends on $M_{\rm BH}$ principally through $\Xi$.  The net result is that $L_{\rm max}$ is greatest for small $M_{\rm BH}$ and large $\mstar$ and least for large $M_{\rm BH}$ and small $\mstar$.

\begin{figure}
	\centering
		\includegraphics[width=8.5cm]{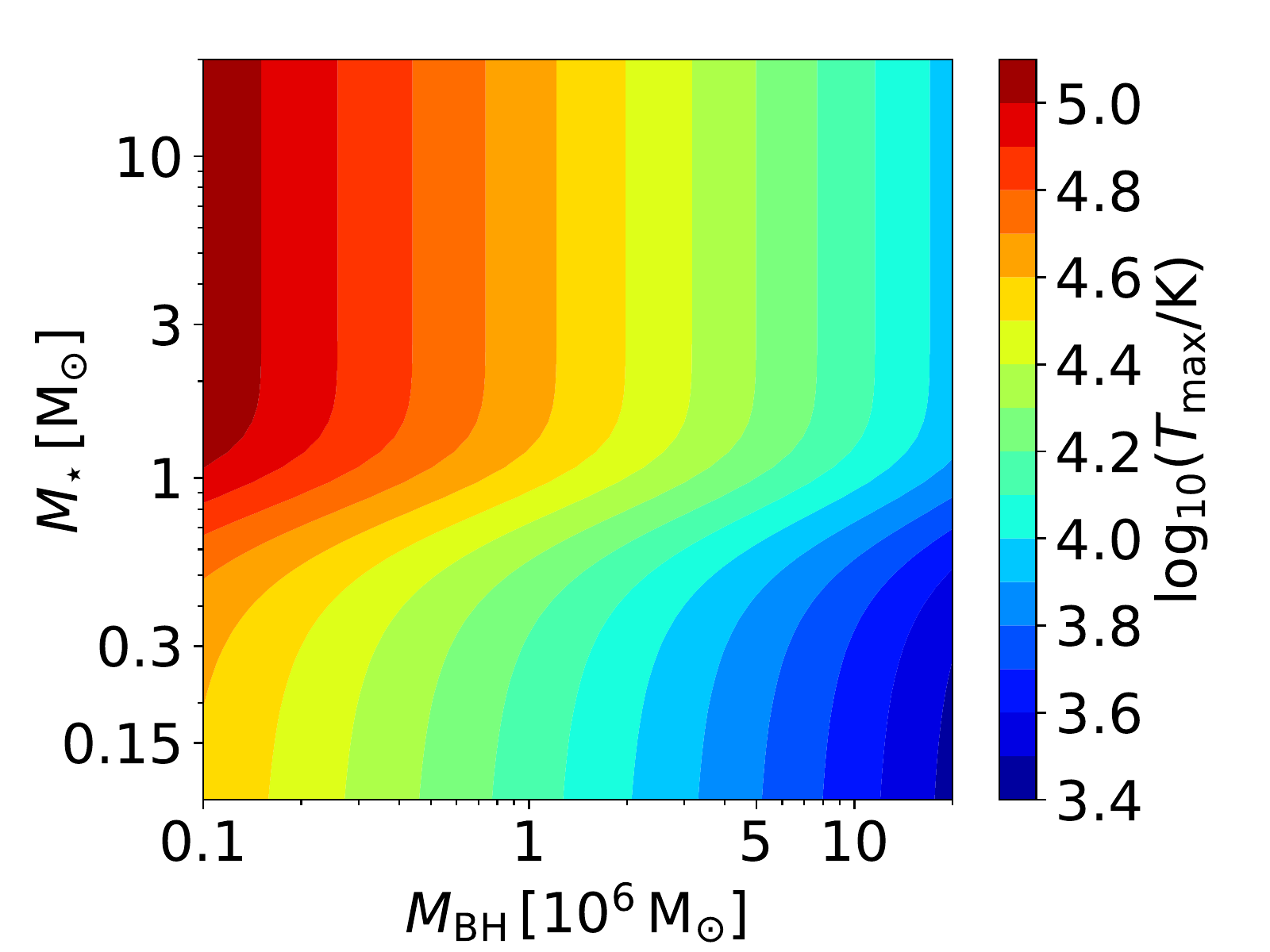}	
	\caption{The maximal blackbody temperature $T_{\rm max}$ (Equation~\ref{eq:Tmax}) in units of K for $c_{1}=1$ and $\Delta\Omega=2\pi$.}
	\label{fig:Tmax}
\end{figure}

This estimate would not hold if the dissipated heat were not radiated promptly.  For example, \citet{Jiang+2016} simulated a collision between two debris streams and found that the energy dissipated in the shock is returned to kinetic energy by adiabatic expansion before many photons can diffuse out.  However, this result depends directly upon the assumption that the streams interact in isolation.  As pointed out by \cite{Piran+2015}, this is not the case in the weakly-circularizing configuration we consider; most of the matter that has fallen back up to this point remains in an orbit with apocenter $\sim a_0$, obstructing such free expansion.

The radiation efficiency therefore depends instead on how the photon diffusion time $t_{\rm diff}$ compares to the accretion inflow time, while its relation to the fallback rate depends on the ratio $t_{\rm diff}/t_0$. The debris near the apocenter is very optically thick: the local optical depth to the midplane is $\tau\sim\kappa (\mstar/2) /[2\pi (c_{1}a_{0})^{2}]\sim 110~c_{1}^{-2}~\mbhs^{-4/3}~\mstar^{5/9}~\Xi^{2}$, where $\kappa=0.34\cm^{2}\g^{-1}$ is the Thomson opacity. The corresponding photon diffusion time is then 
\begin{align}\label{eq:tdiff}
t_{\rm diff} &\sim \frac{\tau h}{c} \sim 1.3 \times 10^{6}\s~\left(\frac{h/r}{0.5}\right)\left(\frac{\Delta\Omega}{2\uppi}\right)^{-1}c_{1}^{-1}~\mbhs^{-2/3}~\mstar^{7/9}~\Xi \ ,
\end{align}
where $h/r$ is the disk aspect ratio.  Thus, we find
\begin{align}\label{eq:ratio_tdiff_t0}
\frac{t_{\rm diff}}{t_{0}}& \sim 0.4~~\left(\frac{h/r}{0.5}\right) c_{1}^{-1}~\mbhs^{-7/6}~\mstar^{4/9}~\Xi^{5/2}\ .
\end{align}
As previously estimated in \citet{Piran+2015}, the photon diffusion time is generally comparable to the matter's orbital period, which is also the characteristic fallback time.  For this reason, it should radiate efficiently, and its lightcurve is related to the fallback rate, but may not reproduce it exactly.

For an emitting region with an effective surface area $\Delta \Omega ~( c_1 a_{0})^{2}$ , where $\Delta \Omega $ is the solid angle, the peak blackbody temperature $T_{\rm max}$ is
\begin{align} 
T_{\rm max} &= \left[ \frac{L_{\rm max}} { \sigma  \Delta \Omega c_{1}^2 a_{0}^2 }\right]^{1/4},\nonumber\\
&=
2.3 \times 10^{4}~\left(\frac{ \Delta \Omega}{2 \pi}\right)^{-1/4} c_{1}^{-3/4} \mbhs^{-3/8}~\Xi^{9/8}~{\rm K},
\label{eq:Tmax}
\end{align}
where $\sigma$ is the Stefan-Boltzmann constant.  
We set $2\uppi$ as the fiducial value of $\Delta\Omega$ because we expect the emission surface to be somewhat flattened, both the top and bottom surfaces radiate, and not all the surface is equally heated.
Figure~\ref{fig:Tmax} depicts  the contour map for $T_{\rm max}$. 
Unlike $L_{\rm max}$, $T_{\rm max}$ depends   predominantly on $\mbh$; its only connection to $\mstar$ is via $\Xi$.

\begin{figure}
	\includegraphics[width=8.5cm]{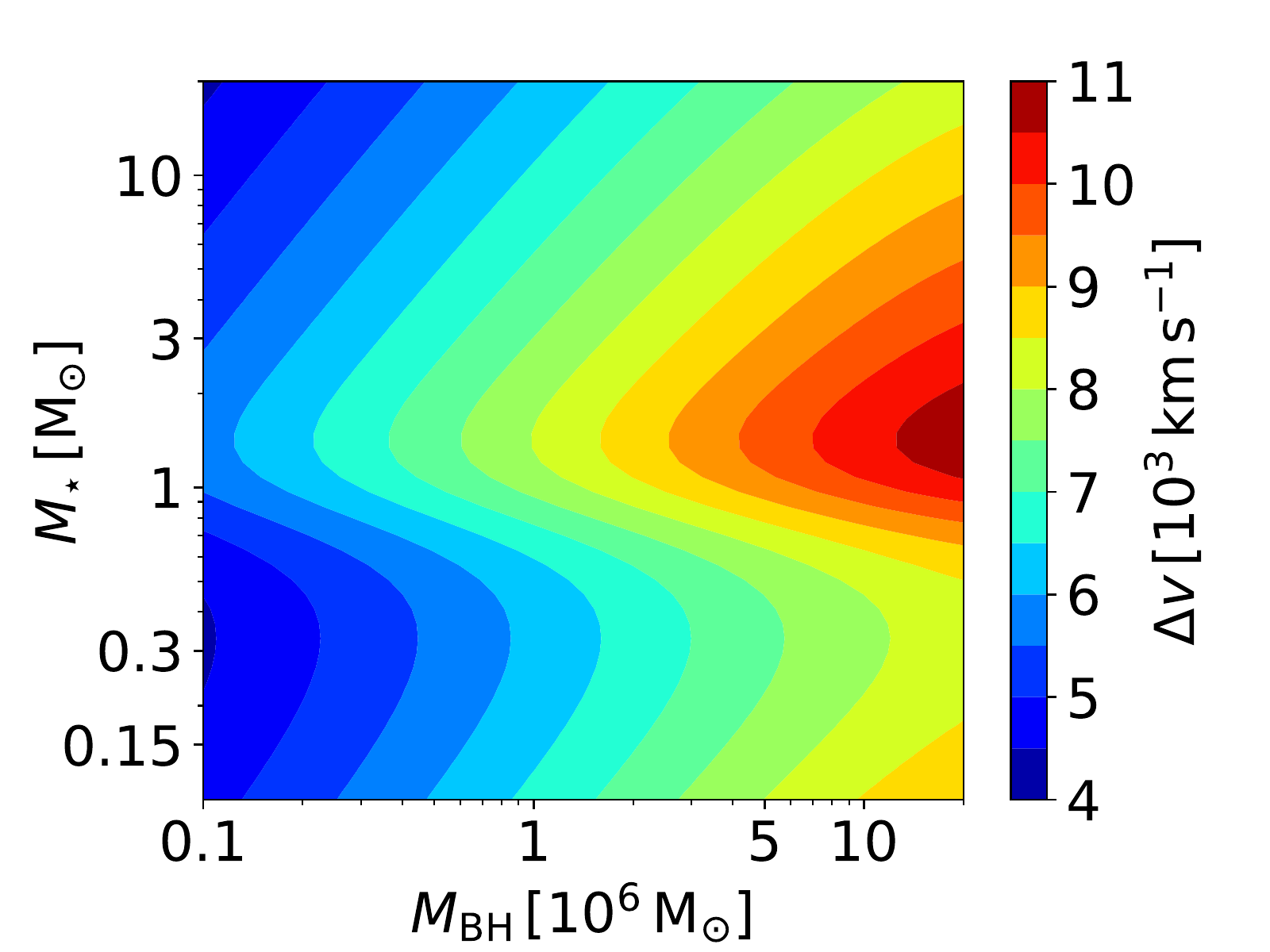}	
	\caption{Contour lines for the line width $\Delta v$ in units of $10^{3}~$km$~\s^{-1}$ for $c_{1}=1$ and $\Delta\Omega=2\uppi$.}
	\label{fig:linewidth}
\end{figure}

This model also predicts the characteristic orbital speed of matter in the accretion flow:
\begin{align} 
\label{eq:deltav}
\Delta v = \big[ \frac{2 \Delta E}{c_{1}}\big]^{1/2}  =  6400~\mbhs^{1/6} \mstar^{-1/9}c_{1}^{-1/2}~\Xi^{1/2}~{\rm km/s}.
 \end{align} 
This dependence is illustrated in Figure~\ref{fig:linewidth}.   As the algebraic relation demonstrates, $\Delta v$ is extremely insensitive to all the parameters.  That it is in the middle of the observed range is encouraging, but its insensitivity to $M_{\rm BH}$ and $\mstar$ makes it not very useful for parameter inference.   Moreover, quantitative matching to {\it observed} line profiles involves the line-of-sight velocity $\Delta v \left(c_1 a_0/r-1\right)^{1/2}\sin i\ f(\varpi)$, where $r$ is the radius of a fluid element from the black hole as it follows an eccentric orbit with apocenter $c_1 a_0$, $i$ is the inclination of the orbital plane to our line-of-sight, and $\varpi$ is the angle between the line of apses and our line-of-sight\footnote{$f(\varpi) = (q\sin\phi \cos\varpi + \sin\varpi)/(1 + q^2\sin^2\phi)^{1/2}$, where $q=2er/[a_0(1-e^2)]$ and the line of apses defines $\phi=0$.}.  We therefore don't use it in our mass estimates. However, it provides a useful approximate consistency check.

\begin{figure*}
		\includegraphics[width=8.5cm]{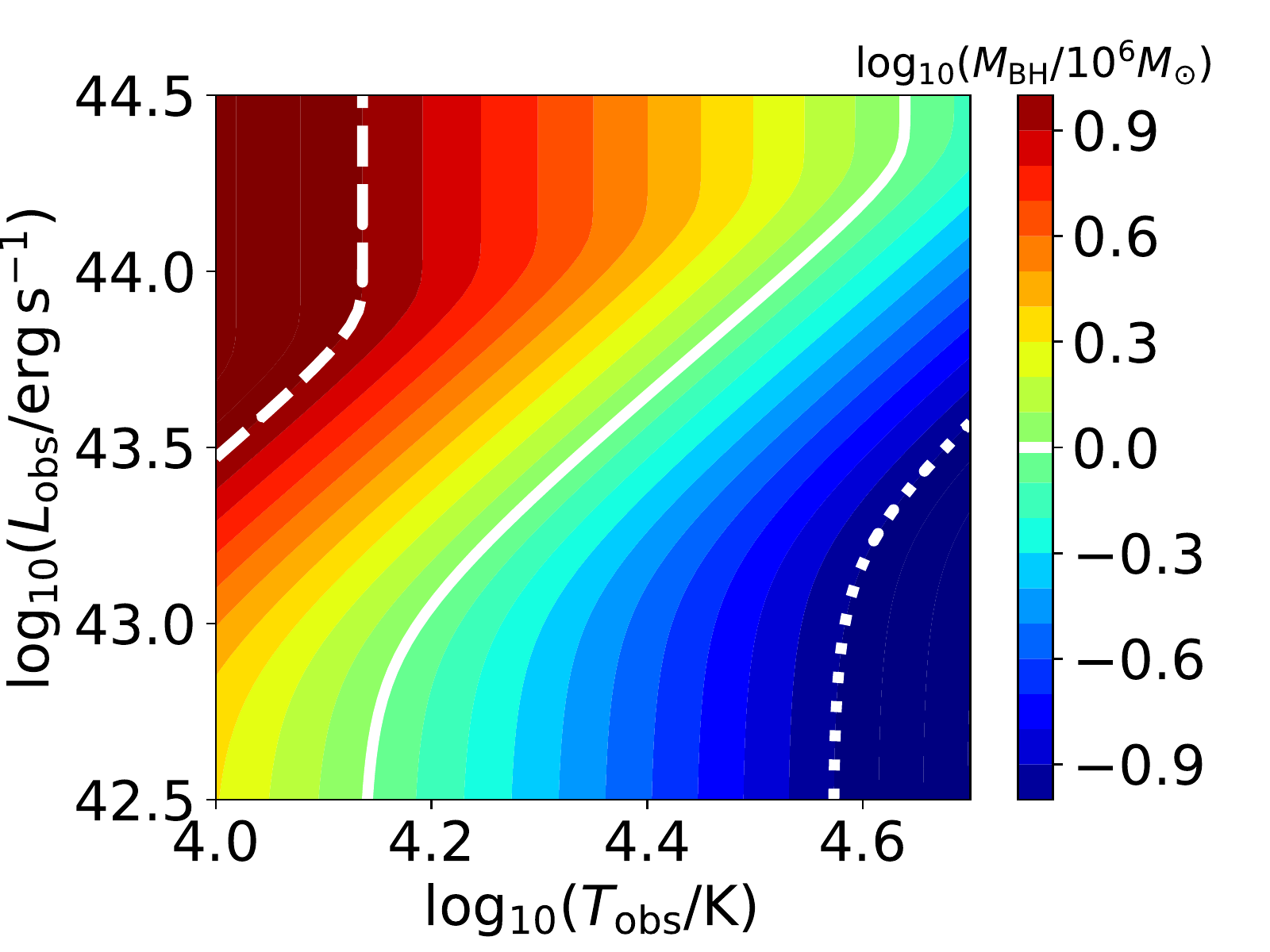}	
		\includegraphics[width=8.5cm]{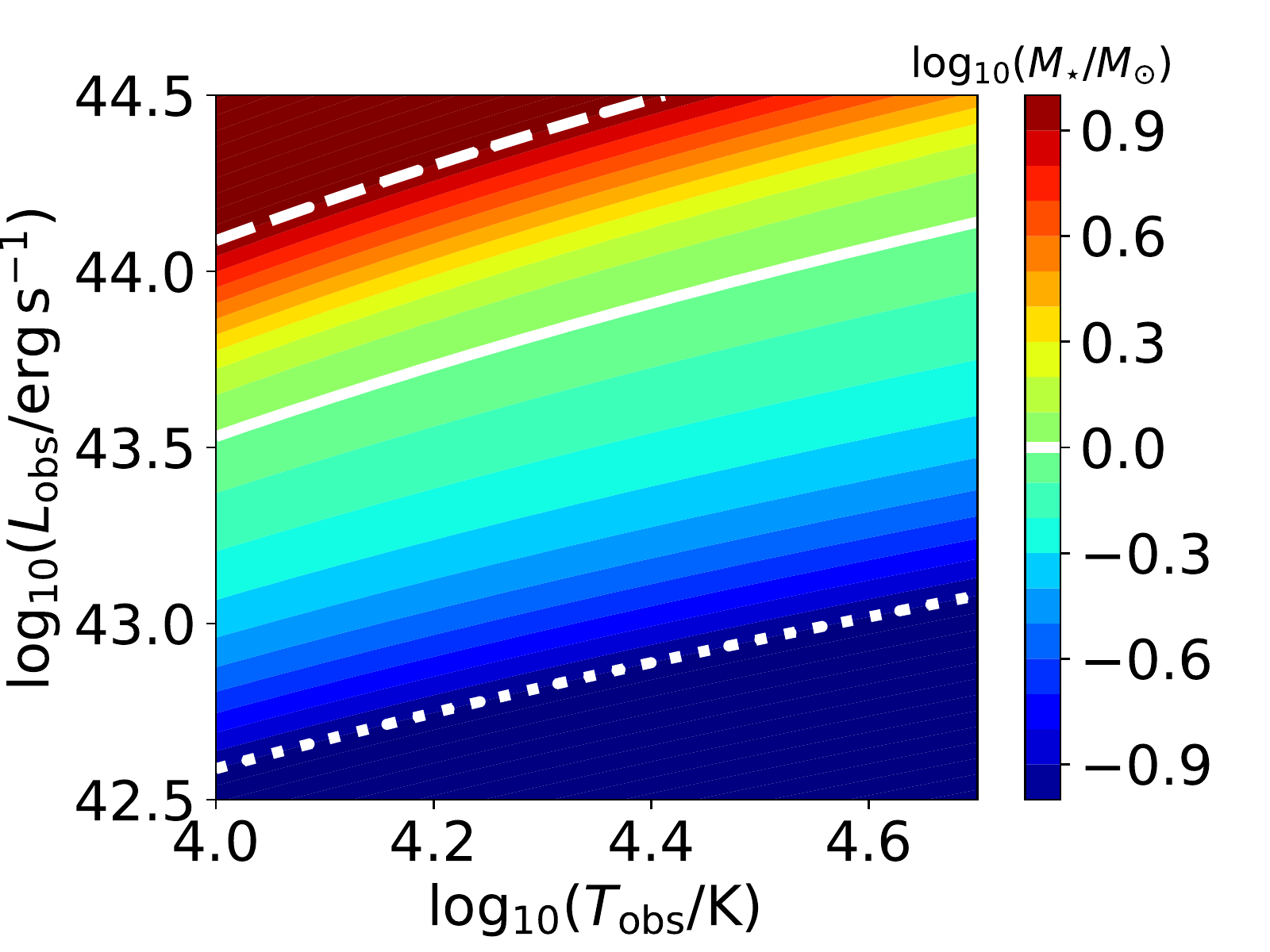}
	\caption{The inferred $\mbh$ and $\mstar$ for $c_{1}=1$ and $\Delta\Omega=2\uppi$  within the ranges of $\Lobs$ and $\Tobs$ for typical UV/optical TDE candidates. The white dashed (dotted) line indicates the upper (lower) bound of the color scale. The white solid line indicates $\mbh = 10^{6}\Msun$ in the \textit{left} panel and $\mstar=1\Msun$ in the \textit{right} panel.}
	\label{fig:mbh_mstar_sol}
\end{figure*}

\section{From observables to parameters} 
\label{sec:from_obs_to_par}

\subsection{Inversion of the model equations}
Consider an event with an observed peak luminosity $\Lobs$ and observed temperature $\Tobs$ at the time of peak luminosity.
Although most of the events discovered so far were found after they reached their peak, recently a number of TDEs have been identified in which the peak was observed \citep[e.g.,][]{nicholl2020,Hinkle+2020b}.
In this case $\Lobs$ and $\Tobs$ correspond to  $L_{\rm max} $ and $T_{\rm max}$\footnote{When $L_{\rm obs} < L_{\rm max}$, $\mstar$ can be underestimated.  Often, $\Tobs$ is almost constant through the event \citep{Hung+2017, Hinkle+2020b}; if so, the sensitivity of $\mstar$ on $\Lobs$ is reduced by $\Xi^{-9/2}$ for $0.5\lesssim\mstar\lesssim1.5$, resulting in $\mstar\propto \Lobs^{0.4-0.7}$. But for other values of $\mstar$, the weak dependence of $\Xi$ on $\mstar$ restores the sensitivity of $\mstar$ to $L_{\rm obs}$ ($\mstar\propto \Lobs^{9/4}$). This issue will become less important in future surveys with short cadences.}.
Inverting Equations \ref{eq:mbh_Tobs} and \ref{eq:mstar_Lobs} we find
the two key equations of our model:
\begin{align}\label{eq:mbh_Tobs}
   \mbhs &= 0.5~T_{\mathrm{obs},4.5}^{-8/3} ~\left(\frac{ \Delta \Omega}{2 \uppi}\right)^{-2/3}~  c_{1}^{-2} ~\Xi^{3},
\end{align}
and 
\begin{align}
\label{eq:mstar_Lobs}
   \mstar &= 5~ L_{\rm obs,44}^{9/4} ~ T_{\mathrm{obs},4.5}^{-1} ~(\frac{ \Delta \Omega}{2 \uppi})^{-1/4} ~c_{1}^{3/2}~\Xi^{-9/2}.
\end{align}
Here, $L_{\rm obs} = 10^{44}~\erg~\s^{-1}~L_{\rm obs,44}$ and $T_{\rm obs}= 30000\Kelvin ~T_{\rm obs,4.5}$.
Note that $\mbh$ is primarily determined by $\Tobs$ (Equation~\ref{eq:mbh_Tobs}) and $\mstar$ by $\Lobs$ (Equation~\ref{eq:mstar_Lobs}). Both are strongly dependent on $\Xi$. 

\begin{figure*}
	\includegraphics[width=8.5cm]{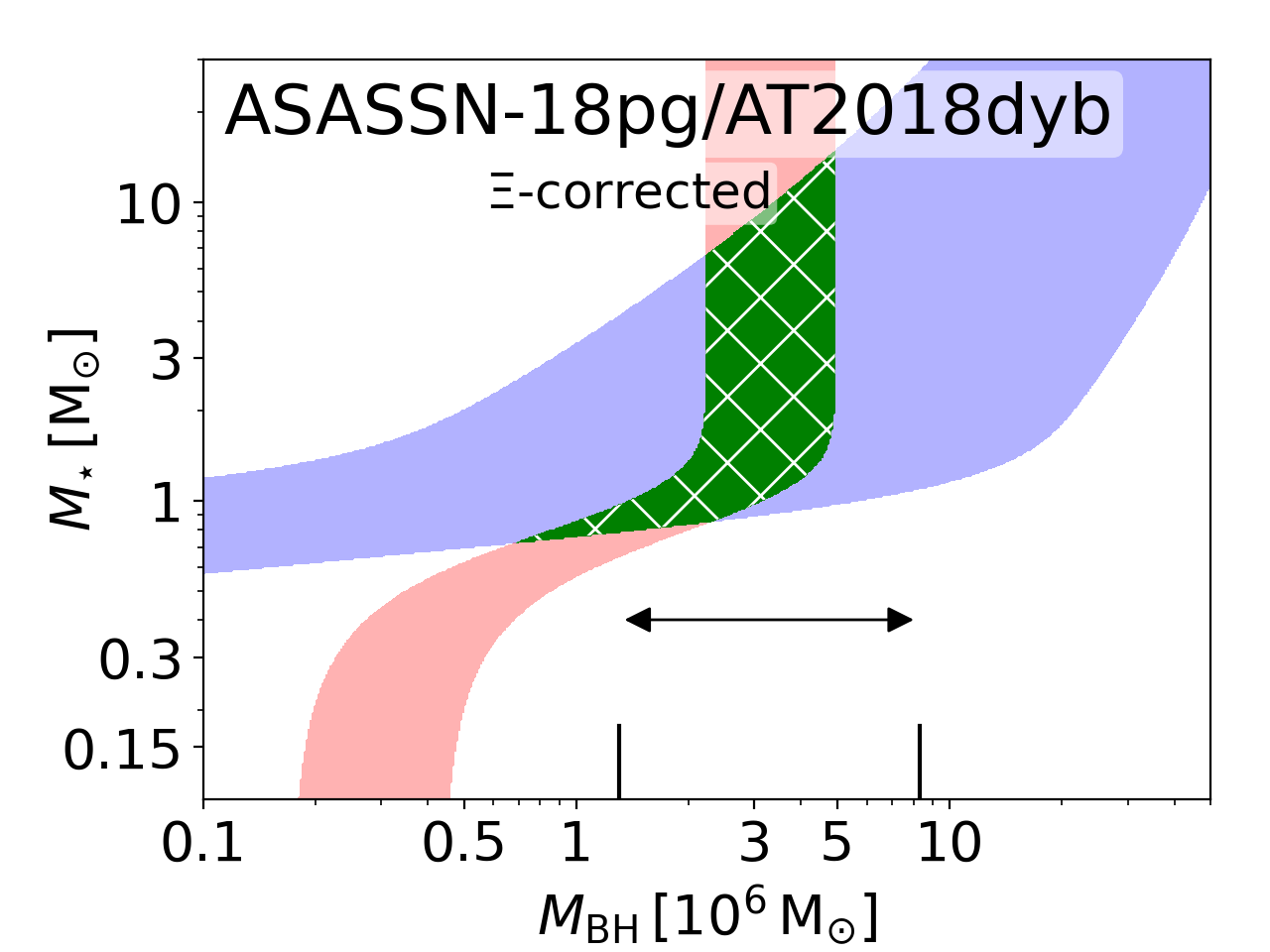}
	\includegraphics[width=8.5cm]{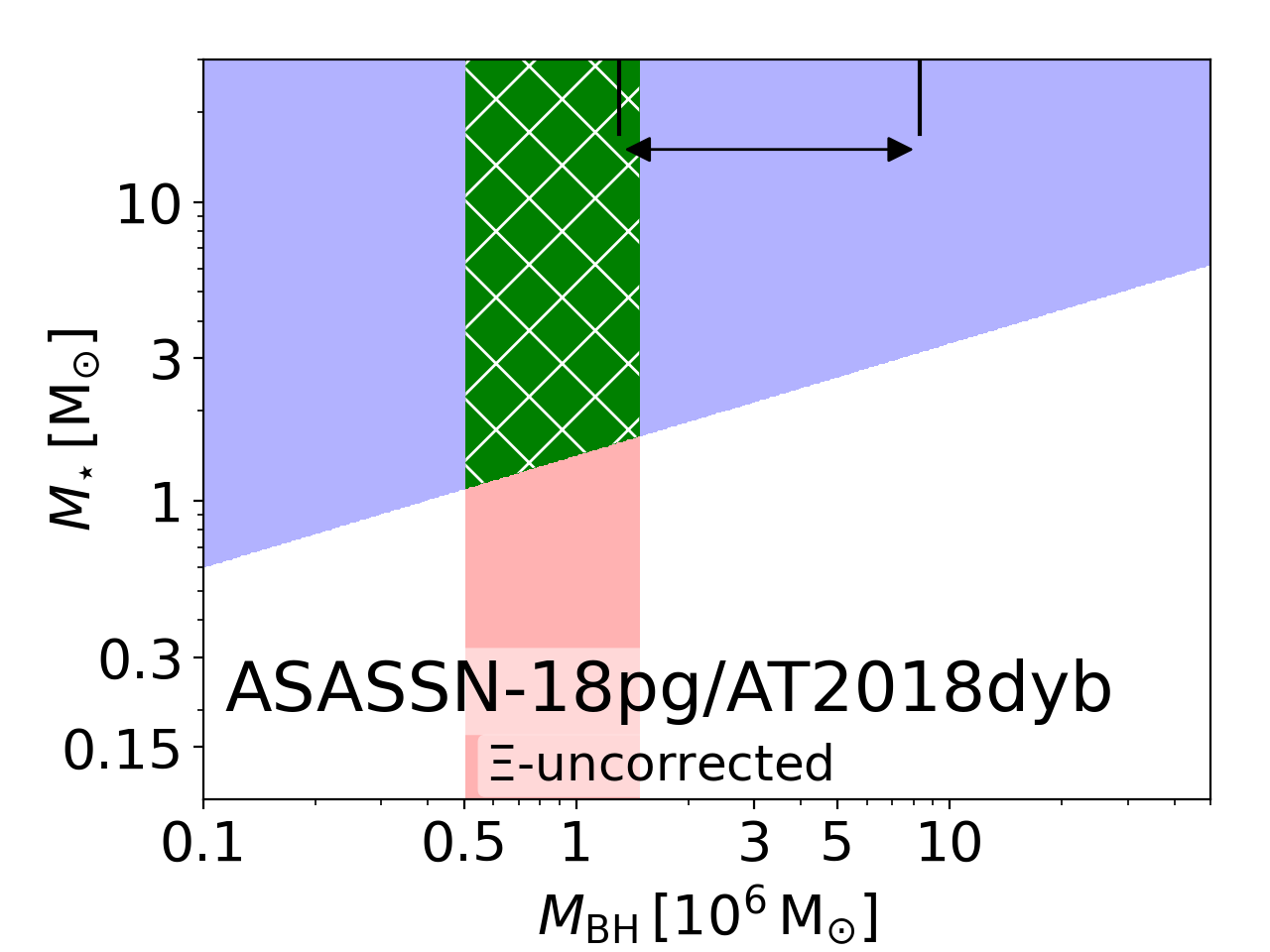}
	\caption{$\mbh$  and $\mstar$ inferred for ASASSN-18pg/AT2018dyb including $\Xi(M_{\star},M_{\rm BH})$ (the \textit{left} panel) and setting $\Xi=1$ (the \textit{right} panel). The red and blue strips correspond to the solution of Equations~\ref{eq:mbh_Tobs} and \ref{eq:mstar_Lobs}, respectively, with $c_{1}=1$ and $\Delta\Omega=2\pi$. The width of each strip is defined by the measurement uncertainty of the observed data. The green X-hatched area where the two strips intersect indicates the permitted range of $\mbh$ and $\mstar$ for the given $\Lobs$ $\Tobs$. The arrows in both panels show the range of $\mbh$ estimated using a bulge-BH correlation \citep{Leloudas+2019}.
	}
	\label{fig:example}
\end{figure*}

\begin{figure}
	\includegraphics[width=8.5cm]{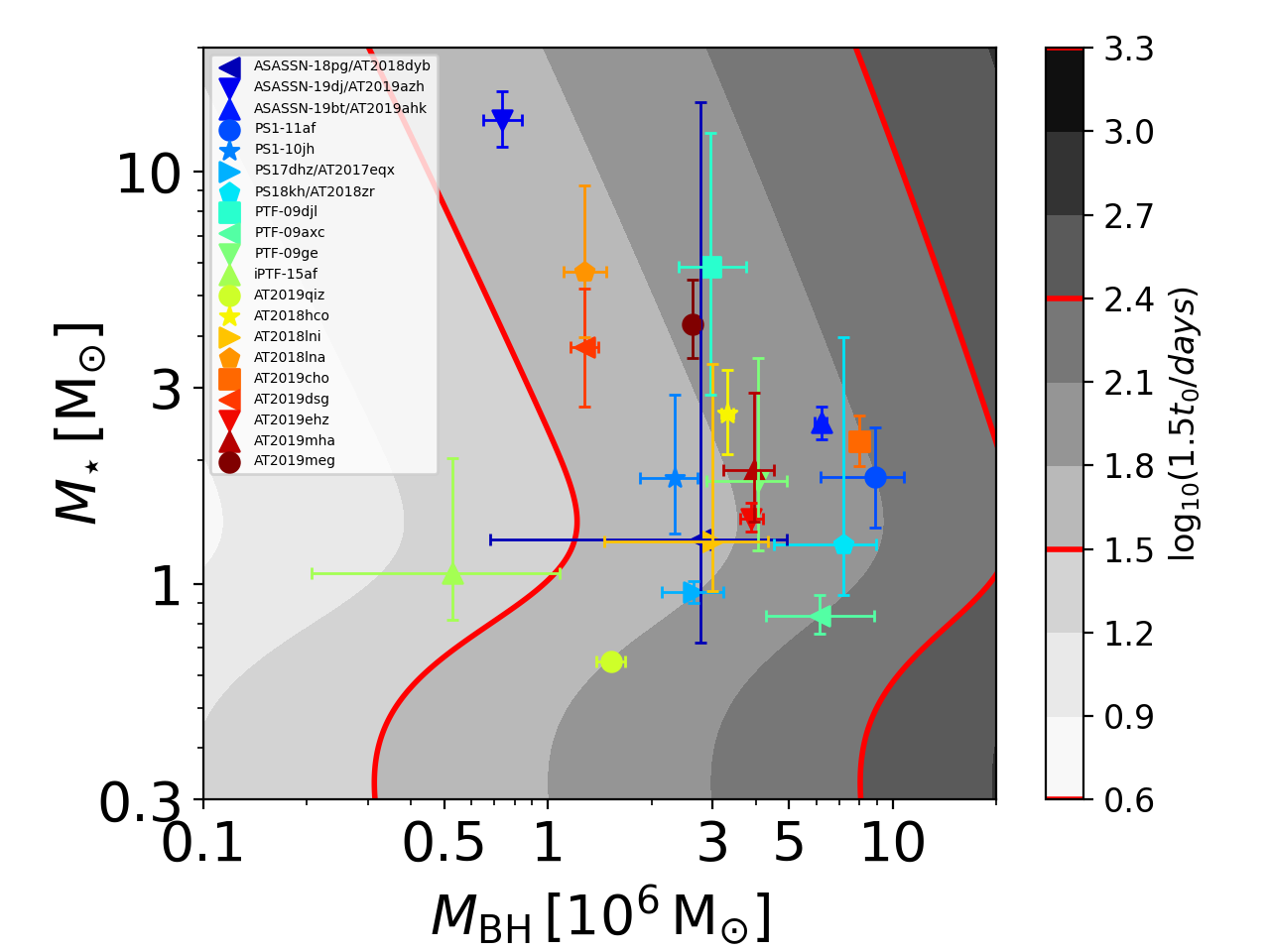}
	\caption{$\mbh$ and $\mstar$ inferred from $\Lobs$ and $\Tobs$ for the a selected TDE sample with $c_{1}=1$ and $\Delta\Omega=2\uppi$. Each filled marker indicates the solutions for given $\Lobs$ and $\Tobs$. The horizontal (vertical) error bar is the range between the extreme values of the inferred masses with the uncertainties of $\Lobs$ ($\Tobs$). The background contour plot show the characteristic time scale ($1.5t_{0}$) for the date at peak luminosity since disruption in units of days.}
	\label{fig:inferredmass}
\end{figure}

Because $\Xi$ depends non-linearly on $\mbh$ and $\mstar$, Equations~\ref{eq:mbh_Tobs} and \ref{eq:mstar_Lobs} must be solved numerically.
We do this either by interpolating within precalculated tables of $L_{\rm max}(\mbh,\mstar)$ and $T_{\rm max}(\mbh,\mstar)$ or by using a 2-dimensional Newton-Raphson method. Solutions of Equations~\ref{eq:mbh_Tobs} and \ref{eq:mstar_Lobs} are shown in Figure~\ref{fig:mbh_mstar_sol} for the ranges of $\Lobs$ and $\Tobs$ relevant for observed TDE events (Table~\ref{tab:tdecandidates}).
The {\sc Python} code implementing our solution is available at \href{https://github.com/taehoryu/TDEmass.git}{https://github.com/taehoryu/TDEmass.git}.

\subsection{An example}
Figure \ref{fig:example} shows an example illustrating both our method and the importance of the $\Xi$ factor: $\mbh$ and $\mstar$ inferred for ASASSN18pg/AT2018dyb \citep{Leloudas+2019}. The red and blue strips demarcate the ranges of the solutions for Equations \ref{eq:mbh_Tobs} ($\Tobs$) and \ref{eq:mstar_Lobs} ($\Lobs$), respectively. The solutions shown in the \textit{right} panel use $\Xi=1$. The inferred black hole mass using $\Xi \neq 1$ is $2.8_{-2.1}^{+2.2}\times10^{6}$ and the inferred stellar mass is $1.3_{-0.6}^{+13.6}$. The errors
 are defined by the range between the extreme values of the inferred mass arising from the uncertainties of $\Lobs$ and $\Tobs$. The inferred black hole mass from our model is consistent with the black hole mass estimated by \citet{Leloudas+2019} using the $M_{\rm BH} - \sigma$ relation of \citet{McConnellMa2013}, $3.3 _{ -2.0 }^{+ 5.0 }\times10^{6}$, which is indicated by an arrow in both panels. 
 Without the $\Xi$ correction, the inferred $\mbh$ is smaller by a factor of 3.5 and it is only marginally consistent with the bulge-inferred black hole mass, while the inferred $\mstar$ is larger by
 a factor that could be as much as $\sim 6$.
In the following section, we will apply our method to a larger sample.

\section{Application to a TDE sample}
\label{sec:results}

To further demonstrate the method, we apply our model to 21 UV/optical TDE candidates for which the date of first observation is more than 10 days before the time of peak luminosity (longer than twice the typical cadence) so that the peak luminosity and temperature can be well-measured.  This set provides enough examples to explore the use of our method on real cases.
Ten of our examples are from \citet{velzen+2020} (AT2019qiz, AT2018hco, AT2018iih, AT2018lni, AT2018lna, AT2019cho, AT2019dsg, AT2019ehz, AT2019mha and AT2019meg).
Eight more are included at least once in the samples collected in \citet{nicholl2020} and \citet{Hinkle+2020b}: ASASSN-18pg/AT2018dyb \citep{Leloudas+2019}, ASASSN-19dj/AT2019azh \citep{Hinkle+2020}, ASASSN-19bt/AT2019ahk \citep{Holoien+2019b}, PS1-11af \citep{Chornock+2013}, PS1-10jh \citep{Gezari+2012}, PS17dhz/AT2017eqx \citep{Nicholl+2019}, PS18kh/AT2018zr \citep{Holoien+2019} and iPTF-15af \citep{Blagorodnova+2019}.  The last three are from \citet{Arcavi+2014}: PTF-09djl, PTF-09axc, PTF-09ge.

Using the published data for $L_{\rm obs}$ and $T_{\rm obs}$ for each case, we infer the black hole mass and stellar mass, as well as the characteristic orbital period $t_0$ they together imply (see Equation~\ref{eq:t0}).
The results for $\mstar$ and $M_{\rm BH}$, also including $t_0$, are shown in Table~\ref{tab:tdecandidates}. 
We find values of $\mbh$ and $\mstar$ for 20 of the 21 events within the expected range.  Although it is encouraging that our model yields plausible parameters for nearly every case, it is not surprising because, when applied to generic values of $M_{\star}$ and $M_{\rm BH}$, our model {\it predicts} values of $\Lobs$ and $\Tobs$ in the middle of the range of observed values, and with relatively weak dependence on $M_{\star}$ and $M_{\rm BH}$.  However, it is also very striking and encouraging that omission of $\Xi$ significantly degrades its performance: if $\Xi$ is ignored, for 7 of the 21 events, the inferred $M_{\star}$ is $\gtrsim 20\Msun$, so large as to make it implausible given the stellar mass distribution.  This fact immediately emphasizes the importance of using careful calculations of $\Delta E$.  In addition, the fact that use of realistic physics improves performance supports the viability of the underlying model. This point is strengthened by the fact that in nearly all these cases, the degree to which $\Xi \neq 1$ in our full solution is almost entirely due to $M_{\star}$, rather than $M_{\rm BH}$; in other words, correct treatment of the $M_{\star}$-dependence of $\Xi$ changes an unreasonable inferred value of $M_{\star}$ to a reasonable one.

AT2018iih is the one case in which an inferred mass appears to be outside the reasonable range: for this object, we find $\mstar = 75$.  However, examination of the discovery paper (see in particular Figures 1 and 11 of \citealt{velzen+2020}) reveals that this event is an outlier with respect to the rest.  In addition to its high luminosity and low temperature, it also has a very slow decay rate, so that its observed total radiated energy is an order of magnitude or more greater than any of the others.  It may possibly be a misidentified different variety of transient.

Finally, we show in Figure~\ref{fig:inferredmass} the inferred $\mbh$ and $\mstar$ for the observed $\Lobs$ and $\Tobs$ superimposed on contours of $1.5~t_{0}$, the expected delay between stellar pericenter passage and peak light. Note that the range of $M_{\star}$ shown excludes AT2018iih.
We find that 2/3 of the events have $1 < \mstar < 3$ and $10^6 < \mbh < 10^{7}$.

On the other hand, we find only two cases with $\mstar < 1$, but six events with $\mstar\gtrsim 3$.
Although our sample size is too small and too heterogeneous to support any statistical analysis of the distribution of $\mbh$ or $\mstar$,  we note that this relatively large representation of massive stars is consistent with two facts about their host galaxies. All six of these events (ASASSN-19dj: \citealt{Hinkle+2020}; PTF09axc, PTF09djl: \citealt{Arcavi+2014}; AT2018lna, AT2019dsg, AT2019meg: \citealt{velzen+2020}) took place in post-starburst galaxies.  Moreover, a remarkable fraction of all known tidal disruptions happened in galaxies with post-starburst stellar populations \citep{Arcavi+2014,French+16,Law-Smith+17,Graur+18}.
Conversely, we may speculate about why we see comparatively few low-mass stars despite the large population of stars in this mass range.
Smaller $\mstar$ leads to less luminous, hence harder to detect, TDEs. It follows that one possible explanation for the paucity of smaller mass stars is that events with $L_{\rm max}$ large enough to be discovered well before the peak are likely to have larger values of $\mstar$ (see the \textit{right} panel of Figure~\ref{fig:mbh_mstar_sol}).

As shown by both Figure~\ref{fig:inferredmass} and Table~\ref{tab:tdecandidates}, the magnitude of the inferred delay time in this sample ranges from $\approx 50$~d to $\approx 120$~d, excluding AT2018iih.  With only one exception in our entire sample of 21 (again excluding AT2018iih), the ratio $t_{\rm diff}/t_0$ lies between $\sim 0.1$ and $\sim 1$.

\begin{table*}
	\centering
\begin{center} 
	\renewcommand{\thetable}{\arabic{table}}
	\caption{Measured properties ($\Lobs$ and $\Tobs$, $M_{\rm BH, bulge}$) for a TDE sample and the inferred properties $\mbh$, $\mstar$ and $t_{0}$ for $c_{1}=1$ and $\Delta\Omega = 2\pi$. The uncertainties of $\mbh$ and $\mstar$ are defined to enclose their extreme values for the given $\Lobs$ and $\Tobs$. The bulge-inferred black hole mass $M_{\rm BH, bulge}$ (estimated using either a $M_{\rm BH}-M_{\rm bulge}$ relation or $M_{\rm BH}-\sigma$ relation) and its uncertainty are taken from the corresponding cited paper whenever possible (see footnotes for details).
	 For ASASSN-19dj/AT2019azh and PTF-09axc, we present two different bulge-inferred black hole masses (see footnotes ${a}$ and ${c}$).}
	\label{tab:tdecandidates}
	\begin{tabular}{c c c c |c c c |c}
	Candidate name & $\Lobs[10^{44}\erg/{\mathrm s}]$ & $\Tobs[10^4\mathrm{K}]$ 
	& $M_{\rm BH, bulge}[10^6~M_{\odot}]$ 
	& $\mbh[10^6~M_{\odot}]$ &  $\mstar[M_{\odot}]$  & $t_{0}$ [days] & Reference\\
		\hline\hline \noalign{\smallskip}
ASASSN-18pg/AT2018dyb &$ 1.1_{ -0.6 }^{+ 1.9 } $& $ 2.5 _{ -0.5 }^{+ 0.5 }$  &$^{i}3.3_{-2.0}^{+5.0}$ (\citetalias{McConnellMa2013}-a)
& $2.8 _{- 2.1 }^{+ 2.2 }$ & $1.3 _{- 0.6 }^{+ 13.6 }$ &  $ 36 _{- 19 }^{+ 38 }$ & \citetalias{Leloudas+2019}
\\ \noalign{\smallskip}
ASASSN-19dj/AT2019azh & $ 6.2 _{ -0.2 }^{+ 0.2 }$& $ 5.0 _{ -0.3 }^{+ 0.3 }$ & $^{k}12_{-4}^{+7}$ (\citetalias{McConnellMa2013}-b),~$^{a}< 4$ (\citetalias{Gultekin2009})
& $ 0.74_{- 0.09 }^{+ 0.09 }$ & $ 13_{- 2 }^{+ 2 }$ & $ 31 _{- 4 }^{+ 5}$ & \citetalias{Hinkle+2020}
\\\noalign{\smallskip}
ASASSN-19bt/AT2019ahk &$ 1.2_{ -0.0 }^{+ 0.0 }$& $ 1.8_{ -0.0 }^{+ 0.0 }$ & $^{k}6.0_{-2.4}^{+4.1}$ (\citetalias{McConnellMa2013}-b)
& $ 6.2 _{- 0.2 }^{+ 0.2 }$ & $ 2.5_{- 0.2 }^{+ 0.2 }$ &  $ 70 _{- 4 }^{+ 4 }$ & \citetalias{Holoien2020}
\\\noalign{\smallskip}
PS1-11af &$ 0.85_{ -0.02 }^{+ 0.02 }$& $ 1.5 _{ -0.2 }^{+ 0.3 }$ & $^{j}8_{-2}^{+2}$ (\citetalias{HaringRix2004})
& $ 8.9 _{- 2.7 }^{+ 1.8 }$ & $ 1.8_{- 0.4 }^{+ 0.6 }$ & $ 84 _{- 22 }^{+ 21 }$ & \citetalias{Chornock+2013}
\\\noalign{\smallskip}
PS1-10jh &$^{c} 1.6 _{ -0.2 }^{+ 0.3 }$ & $^{c} 2.9 _{ -0.2 }^{+ 0.2 }$  & $^{j}4_{-2}^{+2}$ (\citetalias{HaringRix2004}),  $^{i}0.71_{-0.41}^{+1.2}$ (\citetalias{FerrareseFord2005})
& $ 2.3 _{- 0.5}^{+ 0.4 }$ & $ 1.8_{- 0.5 }^{+ 1.1 }$ & $ 33_{- 6 }^{+ 9 }$ & \citetalias{Gezari+2012}
\\\noalign{\smallskip}
PS17dhz/AT2017eqx &$  0.65_{ -0.06 }^{+ 0.05 }$& $ 2.1 _{ -0.1 }^{+ 0.1 } $ &  $^{b,k}6.8_{-2.3}^{+ 3.5}$ (\citetalias{KormendyHo2013})& $ 2.6 _{- 0.5 }^{+ 0.5 }$ & $ 0.96 _{- 0.06 }^{+ 0.05 }$ &  $ 41 _{- 7 }^{+ 9 }$ & \citetalias{Nicholl+2019}
\\\noalign{\smallskip}
PS18kh/AT2018zr &$  0.68_{ -0.20 }^{+ 0.53 }$ & $ 1.5 _{ -0.1 }^{+ 0.1 }  $ & $^{k}7.7_{-3.3}^{+ 5.8}$ (\citetalias{McConnellMa2013}-b)
& $ 7.2 _{- 2.5 }^{+ 1.8 }$ & $ 1.3_{- 0.3 }^{+ 2.7 }$ &  $ 70_{- 21 }^{+ 39 }$& \citetalias{Holoien+2019}
\\\noalign{\smallskip}
PTF-09djl &$^{c}2.5 _{ -0.5 }^{+ 0.7 }$ & $^{c} 2.6 _{ -0.3 }^{+ 0.3 }$ & $ ^{j}3.6_{ -3.0 }^{+10 }$ (\citetalias{HaringRix2004}), $^{i}0.66_{-0.49}^{+1.7}$ (\citetalias{FerrareseFord2005}) 
& $ 3.0 _{- 0.6 }^{+ 0.8 }$ & $ 5.9_{- 3.0 }^{+ 6.4 }$ &  $ 57_{- 18 }^{+ 29 }$  &\citetalias{Arcavi+2014}
\\\noalign{\smallskip}
PTF-09axc &$^{c} 0.31_{ -0.03 }^{+ 0.04 }$ & $^{c} 1.2 _{ -0.1 }^{+ 0.1 }$  & $ ^{j}2.7 _{ -0.6 }^{+ 0.7 }$ (\citetalias{HaringRix2004}), $^{i}0.48_{-0.32}^{+0.97}$ (\citetalias{FerrareseFord2005})
& $ 6.2 _{- 1.7 }^{+ 2.5 }$ & $ 0.84 _{- 0.07}^{+ 0.09 }$ & $ 83_{- 25 }^{+ 37 }$ & \citetalias{Arcavi+2014}
\\\noalign{\smallskip}
PTF-09ge & $ ^{c} 1.3 _{ -0.3 }^{+ 0.3}$ & $^{c} 2.2 _{ -0.2 }^{+ 0.2 }$ & $^{j} 5.7 _{ -1.0 }^{+ 3.0 }$ (\citetalias{HaringRix2004})
& $ 4.1 _{- 1.1 }^{+ 0.9 }$ & $ 1.8_{- 0.6 }^{+ 1.8 }$ &  $ 48 _{- 11 }^{+ 20 }$& \citetalias{Arcavi+2014}
\\\noalign{\smallskip}
iPTF-15af &$ 1.5_{ -0.5 }^{+ 0.8 }$ & $ 4.9 _{ -0.7 }^{+ 0.9 }$ &  $ ^{i}7.6 _{ -4.4 }^{+ 11 }$  (\citetalias{FerrareseFord2005})
&$ 0.53_{- 0.32 }^{+ 0.54 }$ & $ 1.1 _{- 0.3 }^{+ 1.0}$ & $ 14 _{- 7 }^{+ 13 }$	& \citetalias{ Blagorodnova+2019}
\\\noalign{\smallskip}
AT2019qiz &$ 0.29 _{ -0.01 }^{+ 0.01 }$& $ 1.9 _{ -0.0 }^{+ 0.0 } $ &$^{d,j}1.5_{-1.0}^{2.7}$ (\citetalias{Gultekin2009})
& $ 1.5 _{- 0.1 }^{+ 0.1 }$ & $ 0.65 _{- 0.01 }^{+ 0.01 }$ &   $ 43_{- 3 }^{+ 4 }$&\citetalias{velzen+2020}
\\\noalign{\smallskip}
AT2018hco &$ 1.7 _{ -0.1 }^{+ 0.2 }$ & $ 2.5 _{ -0.1 }^{+ 0.1 } $ &$-$
& $ 3.3_{- 0.2 }^{+ 0.1 }$ & $ 2.6 _{- 0.6 }^{+ 0.7 }$ & $ 47_{- 5 }^{+ 5 }$ &\citetalias{velzen+2020}
\\\noalign{\smallskip}
AT2018iih & $ 5.3_{ -0.4 }^{+ 0.4 }$& $ 1.7 _{ -0.0 }^{+ 0.0 }$&$-$ 
& $ 6.8 _{- 0.2}^{+ 0.2}$ & $ 75 _{- 13}^{+ 14 }$ & $ 234 _{- 20 }^{+ 18}$ &\citetalias{velzen+2020}
\\\noalign{\smallskip}
AT2018lni &$1.1 _{ -0.2 }^{+ 0.6}$& $ 2.4 _{ -0.3 }^{+ 0.4 } $&$-$
& $ 3.0 _{- 1.5}^{+ 1.3 }$ & $ 1.3 _{- 0.3 }^{+ 2.1 }$ &  $ 38_{- 14}^{+ 23 }$ &\citetalias{velzen+2020}
\\\noalign{\smallskip}
AT2018lna &$ 3.5 _{ -0.4 }^{+ 0.6 }$& $ 3.9 _{ -0.3 }^{+ 0.3 } $&$-$
& $ 1.3_{- 0.2 }^{+ 0.2}$ & $ 5.7_{- 1.8 }^{+ 3.5 }$ & $ 33 _{- 6 }^{+ 9 }$ &\citetalias{velzen+2020}
\\\noalign{\smallskip}
AT2019cho &$ 1.0 _{ -0.1 }^{+ 0.1 }$& $1.6_{ -0.0 }^{+ 0.0}  $ &$-$
& $ 8.0 _{- 0.3 }^{+ 0.3 }$ & $ 2.2 _{- 0.3}^{+ 0.3 }$ &  $ 82 _{- 6 }^{+ 7 }$ &\citetalias{velzen+2020}
\\\noalign{\smallskip}
AT2019dsg &$ 2.9 _{ -0.3 }^{+ 0.4 }$& $ 3.9 _{ -0.2 }^{+ 0.2 } $ &$-$
& $ 1.3 _{- 0.1}^{+ 0.1 }$ & $ 3.8 _{- 1.0 }^{+ 1.0 }$ & $ 29 _{- 4 }^{+ 5 }$&\citetalias{velzen+2020}
\\\noalign{\smallskip}
AT2019ehz &$ 1.1 _{ -0.0 }^{+ 0.1 }$ & $ 2.2 _{ -0.1 }^{+ 0.1 } $ &$-$
& $ 3.9 _{- 0.3 }^{+ 0.3 }$ & $ 1.4_{- 0.1 }^{+ 0.1 }$ &   $ 45 _{- 2}^{+ 3 }$&\citetalias{velzen+2020}
\\\noalign{\smallskip}
AT2019mha &$ 1.3 _{ -0.1 }^{+ 0.2 }$ & $ 2.2_{ -0.1 }^{+ 0.2 } $ &$-$
& $ 4.0 _{- 0.7 }^{+ 0.6 }$ & $ 1.9 _{- 0.5 }^{+ 1.0 }$ &  $ 48_{- 8 }^{+ 12 }$ &\citetalias{velzen+2020}
\\\noalign{\smallskip}
AT2019meg &$ 2.3 _{ -0.2 }^{+ 0.2 }$ & $ 2.8 _{ -0.1 }^{+ 0.1 } $ &$-$
& $ 2.6 _{- 0.1 }^{+ 0.1 }$ & $ 4.3_{- 0.7 }^{+ 1.1 }$ & $ 47_{- 4 }^{+ 5 }$ &\citetalias{velzen+2020}
\\\noalign{\smallskip}
\noalign{\smallskip}
		\hline\noalign{\smallskip}
	\end{tabular}
	\end{center}
{	\begin{flushleft}	References: 1) \citet{Leloudas+2019}; 2) \citet{Hinkle+2020}; 3) \citet{Holoien+2019b}; 4)  \citet{Chornock+2013}; 5) \citet{Gezari+2012}; 
	6) \citet{Nicholl+2019}; 7) \citet{Holoien+2019};  8) \citet{Arcavi+2014}; 9) \citet{ Blagorodnova+2019} ;10) \citet{velzen+2020}
	\end{flushleft}
	
\begin{flushleft}$M_{\rm BH,bulge}$-References: (I-a) $M_{\rm BH}-\sigma$ relation \citep{McConnellMa2013}; (I-b) $M_{\rm BH}-M_{\rm bulge}$ relation from \citep{McConnellMa2013}; (II) $M_{\rm BH}-\sigma$ relation \citep{Gultekin2009}; (III) $M_{\rm BH}-M_{\rm bulge}$ \citep{HaringRix2004}; (IV) $M_{\rm BH}-M_{\rm bulge}$ \citep{KormendyHo2013}; (V)  $M_{\rm BH}-\sigma$ relation \citep{FerrareseFord2005}
	\end{flushleft}	
	
	 \begin{flushleft}\label{tab:footnotea} $^{a}$ :  \citet{vanVelzen+2019b} find  $M_{\rm BH,bulge}<4\times10^{6}$, using the $\mbh-\sigma$ relation of \citet{Gultekin2009} and a measured upper bound on the bulge dispersion.\end{flushleft}\vspace{-0.14in}
 \begin{flushleft} $^{b}$ : This black hole mass was determined using the black hole mass -- bulge luminosity of \citet{KormendyHo2013}, but applying it to the {\it total} stellar luminosity.\end{flushleft} \vspace{-0.14in}
	  \begin{flushleft}$^{c}$ : We quote $\Lobs$ and $\Tobs$ from Table 3 in \citet{Wevers+2019} since the reference paper provides the black body radius, rather than $\Lobs$ (PTF-09djl, PTF-09axc and PTF09ge, \citealt{Arcavi+2014}) or only the lower limit of $\Lobs$ (PS1-10jh, \citealt{Gezari+2012}). \end{flushleft} \vspace{-0.14in}
	   \begin{flushleft}\label{footnote:d} $^{d}$ : \citet{nicholl2020} estimate three different $M_{\rm BH,bulge}$ using different $M_{\rm BH}-\sigma$ relations for $\sigma=69.7\pm2.3\km~\s^{-1}$, $0.56_{-0.36}^{+1.0}\times10^{6}$ using the $M_{\rm BH}-\sigma$ from \citet{McConnellMa2013}, $3.3_{-1.8}^{+3.9}\times10^{6}$ using that from \citet{KormendyHo2013} and $1.5_{-1.0}^{+2.7}\times10^{6}$ using that from \citet{Gultekin2009}. Although we quote the last $M_{\rm BH,bulge}$ in the table since it is closest to the inferred $M_{\rm BH}$, other two $M_{\rm BH,bulge}$ are also marginally consistent with $M_{\rm BH}$. \end{flushleft} \vspace{-0.14in}
	  \begin{flushleft} $^{i}$ The cited reference \citep{Wevers+2017} defines the uncertainty as the linear sum of the systematic uncertainty from the bulge relation used and the measurement uncertainty. \end{flushleft}\vspace{-0.14in}
	  \begin{flushleft} $^{j}$ We quote the uncertainties from the cited references, but their authors do not clearly define how they were determined.\end{flushleft}\vspace{-0.14in}
	  \begin{flushleft} $^{k}$ The cited references provide only the central value without any uncertainty.  The uncertainty shown is the scatter in the bulge relation used to estimate the central value.\end{flushleft}}
\end{table*}

\begin{figure}
	\includegraphics[width=8.5cm]{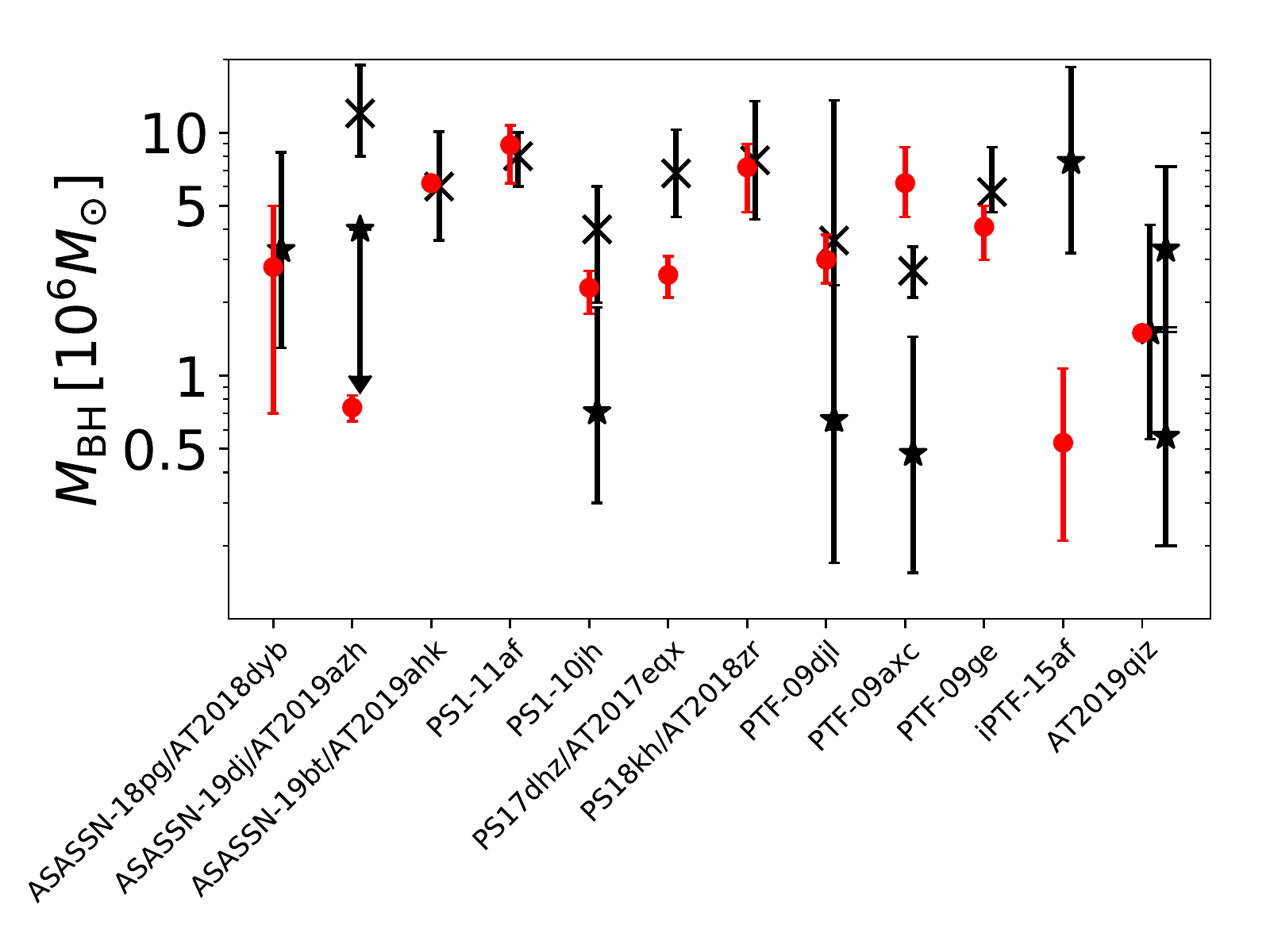}
	\caption{A comparison of $\mbh$ obtained using {\sc TDEmass} (red circles) and $M_{\rm BH, bulge}$ (black symbols) for 12 candidates for which $M_{\rm BH, bulge}$ is available (See Table \ref{tab:tdecandidates} for  references). 
	 For five candidates (ASASSN19dj/AT2019azh, PS1-10jh, PTF-09djl, PTF-09axc and AT2019qiz), we show more than one $M_{\rm BH, bulge}$ estimate taken from published papers, distinguished by different symbols ( crosses : $M_{\rm BH}-M_{\rm bulge}$ relation and stars : $M_{\rm BH}-\sigma$ relation ). The downward arrow attached to one of the two $M_{\rm BH, bulge}$ estimates for ASASSN19dj/AT2019azh indicates an upper bound.}
	\label{fig:mbh_mbh_bulge}
\end{figure}

\section{Discussion}\label{sec:discussion}

\subsection{Comparing our $M_{\rm BH}$ with bulge-inferred black hole mass}

We have just introduced a new way in which central black hole masses can be inferred from TDE observations.  However, central black hole masses can also be estimated from stellar bulge properties, providing a measurement of $\mbh$ independent of any number derived from TDE properties. Here, we define the bulge-inferred black hole mass $M_{\rm BH,bulge}$ as the black hole mass estimated using either a $\mbh-M_{\rm bulge}$ relation or a $\mbh-\sigma$ relation. In Figure~\ref{fig:example}, we showed a case in which our estimate of $M_{\rm BH}$ and that of a bulge-based method coincide quite closely.  In fact, as shown in Figure~\ref{fig:mbh_mbh_bulge}, for 9 of the 12 cases in our sample for which bulge data is available, our inferred $M_{\rm BH}$ is consistent with at least one bulge-based estimate of the black hole mass in that galaxy, and sometimes with more than one  (see Table~\ref{tab:tdecandidates} and Figure~\ref{fig:mbh_mbh_bulge}).  We find this agreement encouraging.

Regrettably, the encouragement we take from this consistency is limited by the difficulties of applying bulge-based estimates. 
First, as has been well-known for a while \citep{Wevers+2017}
and is evident in Figure~\ref{fig:mbh_mbh_bulge}, black hole-bulge correlations can differ substantially. These contrasts are particularly great for the black hole mass range of greatest interest in the TDE context, $M_{\rm BH} < 10^7$, because 
these correlations have been determined primarily by galaxies hosting black holes 1--2 orders of magnitude larger, and the relatively small number of low-mass cases in these samples do not adequately constrain the correlation in this mass range. Three of our cases illustrate this challenge. In one case (ASASSN-19dj/AT2019azh), our inference is consistent with one bulge-based estimate, but not with another. In another (iPTF-15af), our model predicts a value considerably smaller than estimated on the basis of any version of the bulge dispersion correlation. However, as shown by \citet{Xiao+2011} and \citet{Baldassare+2020}, when the dispersion correlations indicate $M_{\rm BH} \lesssim 10^7$, the black hole mass found by emission line widths (when there is an AGN) is generally smaller by factors of several and in some instances is more than an order of magnitude smaller.  In a third (PTF-09axc), different bulge correlations yield $M_{\rm BH}$ estimates differing by a factor $\simeq 5$, but the larger of the two is about a factor of 2 smaller than our inferred value, inconsistent by $\sim 2\sigma$.

Another difficulty is illustrated by a different discrepant case, PS17dhz/AT2017eqx: it can be difficult to resolve the host galaxy well enough to measure the bulge properties.  In this case, the published black hole mass estimate \citep{Nicholl+2019} was made assuming the total stellar mass is the bulge mass; this assumption may explain why our estimate is factor $\sim 2 - 3$ smaller than the ``bulge"-based estimate.

Lastly, uncertainties in these estimates are often quoted as the amount resulting from measurement error in the bulge dispersion or stellar mass. 
However, there is intrinsic scatter in all the correlations: e.g., 0.2-0.5 dex for the \cite{McConnellMa2013} relation and 0.6-0.8 dex for that of \cite{FerrareseFord2005}.  This, too, contributes to the uncertainty.

\subsection{The parameters $c_{1}$ and $\Delta \Omega$}

Our model includes two unspecified parameters, $c_{1}$ and $\Delta\Omega$. 
As seen in Equations~\ref{eq:mbh_Tobs} and \ref{eq:mstar_Lobs}, for fixed $\Lobs$ and $\Tobs$, larger $\Delta\Omega$ results in smaller $\mbh$ and $\mstar$, but the dependence is weak. 
The sensitivity of both $\mstar$ and $\mbh$ to $c_{1}$ is stronger.
For fixed $\Lobs$ and $\Tobs$, $d\log  \mbh / d\log c_{1} \simeq -(1.2-2)$ and $d\log \mstar/ d\log c_{1}  \simeq 0.8-1.5$. 

In this work, we assume $c_{1}=1$ and $\Delta \Omega=2\uppi$ at peak luminosity for all of our TDE sample. Although both assumed values are likely correct to within factors of a few, it will be important to determine both, especially $c_{1}$, more accurately, including any possible dependence on $\mbh$ and $\mstar$
or perhaps on the stellar pericenter $r_{\rm p}$.  \citet{Tejeda+2017} and \citet{Gafton2019} have shown that $\Delta E$ varies only weakly with $r_{\rm p}$ for most pericenters inside the physical tidal radius, but acquires greater sensitivity to $r_{\rm p}$ when it is in the highly-relativistic region, so it could, in principle, influence these two parameters.
If an independent estimate of either $c_1$ or $\Delta\Omega$ becomes available (whether using some new observational constraint or incorporating results from full numerical simulations), it would be possible to constrain these potential dependences. Without such an estimate we recommend keeping them fixed.

\subsection{Characteristic time scale $t_{0}$}
Our model implies  that the peak luminosity should be observed when the tightly bound debris reach apocenter a second time, $\simeq 1.5~t_{0}$ after the star's pericenter passage. This is slightly later than the $\simeq t_0$ delay if a compact accretion disk forms when the debris first return to pericenter.
It follows that
\begin{align}\label{eq:t_constraint}
t_{\rm obs}\gtrsim 1.5~ t_{0}, 
\end{align}
where $t_{\rm obs}$ is the time from the beginning of the disruption event (stellar pericenter passage) to the time of peak light.

This constraint is usable only when it is possible to identify the time at which the disruption began, even though the flare doesn't begin until well after that moment. To accomplish this, one might consider fitting the post-peak light curve assuming the conventional $t^{-5/3}$ power-law, where $t=0$ is the disruption time \citep[e.g.,][]{Miller+2015}. However, this method can be problematic.  There is always the question of the relationship between the mass fallback rate as a function of time and the light curve.  In addition, now that numerous TDES have been observed, it has become clear that their light curves exhibit considerable diversity beyond $t^{-5/3}$: for example, exponentials are better fits to the first few months of ASASSN-14li \citep{Holoien+2015}, ASASSN-14ae \citep{Holoien+2014} and iPTF-16fnl \citep{Blagorodnova+2017}.

An alternative way to identify the disruption moment is to use the radio emission that accompanies some TDEs. \citet{Krolik+2016} found that the radio emission region in ASASSN-14li grew at a constant speed quite close to the propagation speed of the fastest-moving unbound ejecta. By tracing the size of the radio emitting region backwards in time, they inferred the time when the disruption took place, finding it to be $\simeq 70$~days before the TDE was discovered.
We can compare this estimate with the magnitude of $t_0$ derived from the model of this paper, using $L_{\rm obs}$ and $T_{\rm obs}$ rather than lightcurve analysis.  From our new method, we infer $\mbh = 0.52 _{- 0.25 }^{+ 0.54 }\times10^{6}$ and $ \mstar = 0.75 _{- 0.11 }^{+ 0.14 }$ for the event,  giving $t_{0}=20_{- 5 }^{+ 6  }$ days.   Thus, it appears that ASASSN-14li was discovered $\simeq 3.5t_0$ after disruption. 
While longer than expected as there were no observations of this source during this priod  this is not inconsistent.
It is also noteworthy that ASASSN-14li is not a good candidate for light curve-fitting with any simple analytic form because both optical and X-ray luminosities were nearly constant for the first $\simeq 30$~days, and only then began to decline.

When information on the moment of disruption is missing, but the flare has been followed from well before the peak, the time constraint (Equation~\ref{eq:t_constraint}) can be interpreted as a bound on $t_0$: the time from first observation to peak light should be less than $1.5t_{0}$.  Indeed, in the sample presented in this paper, for all our examples $1.5t_0$ is longer than the time from discovery to peak. 
This ratio is $\simeq 2 - 10$ except for one case, iPTF-15af. In this instance, although $1.5t_0$ is only $\simeq 2/3$ the discovery to peak time, the range of masses permitted by the uncertainties in $L_{\rm obs}$ and $T_{\rm obs}$ is such that  $1.5 t_0 \simeq 40$~d is only $1\sigma$ from the central value.

\subsection{Outliers and exceptions}

There can be cases in which our model doesn't apply. This could be because the disruption was only partial or the pericenter was small and the debris circularized rapidly.  In such cases, application of our model may yield values of $\mstar$ or $M_{\rm BH}$ outside its range of validity.

\subsubsection{Partial disruptions}

A significant fraction of all TDEs result in only partial disruption \citep{Krolik+2020}.
In partial disruptions, $\Lobs$ is suppressed by a factor comparable to the ratio of the debris mass to the mass of the star before being disrupted. 
Partial disruptions also differ from total disruptions in the shape of their energy distributions: full disruptions generally have nearly-flat distributions from $-\Delta E$ to $+\Delta E$, while partial distributions create less debris mass with $|E| < \Delta E$ \citep{Guillochon+2013,Goicovic+2019,Ryu3+2020}.  This contrast leads to partial disruptions having more steeply declining mass fallback rates post-peak, and therefore possibly steeper lightcurves than full disruptions.  On the other hand, because $\Delta E$ for partial disruptions resulting in significant mass loss is almost the same as for total disruptions, $t_0$ is little changed.  Because the basic mechanics of apocenter shocks would still operate, their timescales to reach peak should resemble those of full disruptions, while reaching a lower luminosity and then likely declining faster.

A few candidates have been discovered showing hints of these effects, 
e.g., AT2019qiz \citep{nicholl2020} and iPTF-16fnl \citep{Blagorodnova+2017}; they might be partial disruption events. In fact, \citet{Hinkle+2020b} found that in a sample of 21 UV/optical candidates with well-characterized post-peak light curves, 
less luminous TDEs tended to have steeper slopes post-peak than more luminous TDEs.

\subsubsection{Higher $M_{\rm BH}$ and circularization}\label{subsub:highermbh}

In our model, the main source of the observed bolometric luminosity is the heat dissipated by shocks near apocenter.  However, for large $\mbh$ ($\gtrsim (5-10)\times10^{6}$: \citealt{Ryu1+2020}), the tidal radius, when measured in gravitational radii, becomes small, strengthening all relativistic effects, and in particular, apsidal precession. A fraction of events at smaller $M_{\rm BH}$ may involve similarly small pericenters. In this regime, dissipation of the orbital energy into heat takes place in shocks closer to the black hole, on a radial scale closer to the tidal distance $r_{\rm t}$, so that more energy can be dissipated in the shocks, and accretion may proceed more rapidly.  Such a situation also implies considerably higher optical depth, and therefore  time-dependent radiation transfer leading to slower radiation losses.  The degree to which our model may apply in these conditions is unclear. On the other hand, such events should be rare because a large fraction of all passages by stars this close to the black hole result in direct capture by the black hole \citep{Krolik+2020}.

\subsubsection{Examples}\label{subsub:examples}

AT2018iih is a good example of how implausible inferred parameters can signal possible inapplicability or our model: in this case our analysis yielded a nominal $\mstar=75$.  As mentioned earlier, other properties of that event (i.e., beyond $L_{\rm obs}$ and $T_{\rm obs}$) are so different from those of other TDEs  that it may not be a TDE at all.

To a lesser extent, it is possible that ASASSN-19dj/AT2019azh, with its inferred stellar mass of $13$ combined with a rather low SMBH mass, is also an outlier. But in this case it may be a TDE with different characteristics. In particular, in our model its high luminosity (the greatest in our sample), leads to a large stellar mass; if this were instead a case with a small pericenter, more energy would have been dissipated, through either more efficient stream shocks or accretion.  If the resulting structure has a photosphere on a scale $\sim a_0$ (plausible because the energy per unit mass doesn't change as a result of dissipation), effects like these may explain the high temperature and luminosity in this case.

\subsection{Contrast in approach with other methods}
\label{sec:mosfit}

{\sc TDEmass} is a tool to infer $\mbh$ and $\mstar$ for TDEs with optical/UV data.  It differs in many ways from the  method most commonly used hitherto, {\sc Mosfit} with the TDE module \citep{Mockler+2019}. 
The contrast begins with their physical foundation.
{\sc Mosfit} is built upon the assumption that the debris joins an accretion disk of radius $\simeq2~r_{\rm t}$ immediately upon fallback. Soft X-rays are radiated from this disk (after an optional ``viscous" delay) with relativistic radiative efficiency, and the entire X-ray luminosity is reradiated to the optical/UV band by a posited distant reprocessing shell. To apply this model to a specific event demands 6 free parameters in addition to $M_*$ and $M_{\rm BH}$.
By contrast, {\sc TDEmass} 
ascribes the optical/UV emission to the apocenter shocks inevitably caused by small-angle apsidal precession. Using two order-unity parameters held fixed for all cases, it directly determines $M_{\star}$ and $M_{\rm BH}$ for each TDE from $L_{\rm obs}$ and $T_{\rm obs}$ as measured in that event. Thus, the two methods are based on strongly contrasting dynamical pictures and use observational data very differently.
Not surprisingly, they can typically lead to different results. 

More recently \citet{Wen+2020} suggested a different fitting method based on the X-ray spectrum. This model posits a slim disk to produce the X-rays. Like {\sc Mosfit} it, too, supposes quick formation of thin disk, but differs from {\sc Mosfit} in two ways. It ties the light curve to the long-term evolution of the disk rather than to the mass fallback rate, and it attempts to infer the black hole spin rather than the disrupted star mass.

\citet{Zhou+2020} proposed a method to infer $\mbh$ and $\mstar$ based on the elliptical accretion disk model of \citet{Liu+2017}. 
Although both their model and ours posit an elliptical accretion flow, they choose different heating mechanisms. Rather than energy dissipation by shocks, optical/UV luminosity in their model is powered by dissipation associated with accretion from the matter's initial orbit to a smaller orbit (still highly eccentric) whose pericenter permits direct plunge into the black hole.
Their model also differs procedurally: they use the peak luminosity and the total radiated energy (estimated by assuming a post-peak lightcurve $\propto t^{-5/3}$) to infer $\mbh$ and $\mstar$, rather than $\Lobs$ and $\Tobs$.

Comparing cases treated by ourselves and \cite{Mockler+2019} or by ourselves and \cite{Zhou+2020}, we find that both of the other methods yield smaller values of $M_{\star}$: all four of the overlapping cases in \cite{Mockler+2019} have $M_{\star}\sim 0.1$, while those shared with \cite{Zhou+2020} lie in the range $M_{\star}\sim 0.1 - 1$.   
The values of $M_{\rm BH}$ produced by the \cite{Mockler+2019} method are more similar to ours, but those given by the \cite{Zhou+2020} approach are generally a factor of several smaller.  That these three methods lead to different parameter inferences is unsurprising given their very different assumptions about how the light is generated.

\section{Conclusions and summary} \label{sec:conclusions}

We present,  {\sc TDEmass}, a new method to infer black hole mass and stellar mass for optical/UV TDEs 
(available at \href{https://github.com/taehoryu/TDEmass.git}{https://github.com/taehoryu/TDEmass.git}). The method
uses the UV/optical luminosity and the black body temperature at the time of peak luminosity. It is based on the model by \citet{Piran+2015}, in which the optical/UV luminosity arises due to shocks dissipating orbital energy into heat at a distance $\sim a_{0}$. A critical element of this new method is that it incorporates a correction factor, $\Xi$ \citep{Ryu1+2020}, that quantitatively adjusts the classical order of magnitude estimate of the debris energy spread.
It is important to point out that since only spectral data at peak are used, no assumptions for the temporal trends of light curves and their relations to mass fallback rates are made.

Applying our model to 21 examples of TDEs with light curves observed well before peak, we find black hole  and stellar masses within the range $\mstar\simeq 0.65-13$ and $\mbh\simeq 0.5\times10^{5}-10^{7}$ for 20 of the 21 (see Table~\ref{tab:tdecandidates}). The one exception, AT2018iih, is sufficiently an outlier to the others in many respects that it may be a different sort of transient.
For nearly all cases with a black hole mass estimated from bulge properties, our inferred $M_{\rm BH}$ is consistent with the bulge inference, but this is a weak consistency because there can be significant systematic uncertainties in the bulge inference.

{\sc TDEmass} is completely direct---our two inferred quantities may be computed in terms of analytic expressions involving $L_{\rm obs}$ and $T_{\rm obs}$. No other parameters are tuned to fit the data of individual objects. The theory underlying it does, however, possess two order-unity parameters whose proper calculation as functions of $M_{\star}$ and $M_{\rm BH}$ demands numerical simulation---for numerous pairs of $M_{\star}$ and $M_{\rm BH}$---of entire TDEs, from disruption to energy dissipation and light emission. 
The black hole mass is particularly sensitive to one of these parameters ($c_1$); the generally good agreement between our inferences setting $c_1=1$ and values estimated from bulge properties suggests that $c_1 \simeq 1$, as we have chosen here, may not be a bad approximation.

Interestingly, the black hole mass in this model is almost solely determined by the optical/UV effective temperature observed at peak luminosity (see Equations~\ref{eq:Tmax} and \ref{eq:mbh_Tobs} and Figure~\ref{fig:mbh_mstar_sol}), or alternatively, the blackbody radius (see Equation~\ref{eq:a0} and Figure~\ref{fig:a0}).  The one-to-one functional relation between $T_{\rm obs}$ and $M_{\rm BH}$ is modified only through the $\mstar$ dependence of $\Xi$. Thus, $T_{\rm obs}$ all by itself can provide a first rough estimate of $\mbh$.  On the other hand, the stellar mass $\mstar$ depends mostly on the peak luminosity (see Figure~\ref{fig:mbh_mstar_sol}). Therefore,  within this model these two quantities are determined almost independently; they are coupled primarily in the parameter range for which $\Xi$ changes rapidly as a function of $\mstar$ (see Figure \ref{fig:xi}).

The close connection between tidal disruption dynamics and light production in its underlying model makes {\sc TDEmass}  an attractive tool for physical parameter inference in these dramatic events. 
Because it is founded upon a clear physical model, its applicability has well-defined limits; in particular, it is best-justified for total disruptions whose stellar pericenter is large enough that circularization is slow. 
Cases outside this range, whether it is because they are partial disruptions, the black hole mass is too large, the pericenter is too small, or they are not TDEs at all can be readily recognized.
Lastly, in the future, when TDE samples with clear selection criteria become available, this method could be used to infer population properties of both supermassive black holes and stars in galactic nuclei.

\section*{Acknowledgements}
We thank Iair Arcavi and Nicholas Stone for helpful comments. We are grateful to the anonymous referee for some useful comments. This research was partially supported by an advanced ERC grant TReX and by NSF grant AST-1715032.

\bibliographystyle{mnras}

\end{document}